\documentclass[useAMS,usenatbib]{mn2e}

\usepackage{xspace}
\usepackage{graphicx}
\bibpunct{(}{)}{;}{a}{}{,}
\def\lya{\mbox{Ly$\alpha$}\xspace}
\def\lyb{\mbox{Ly$\beta$}\xspace}
\def\mic{\mbox{$\mu$m}\xspace}
\def\kms{\mbox{$\mathrm{km\,s^{-1}}$}\xspace}
\newcommand{\ebv}{\ifmmode E_{\rm B-V} \else $E_{\rm B-V}$ \fi}
\def\msun{\ifmmode M_{\odot} \else M$_{\odot}$\fi}
\def\mstar{\ifmmode M_{\star} \else M$_{\star}$\fi}
\def\msunyr{\ifmmode M_{\odot} {\rm yr}^{-1} \else M$_{\odot}$ yr$^{-1}$\fi}
\def\zsun{\ifmmode Z_{\odot} \else Z$_{\odot}$\fi}
\def\lsun{\ifmmode L_{\odot} \else L$_{\odot}$\fi}

\title[The brightest \lya emitting galaxies]{The properties of the brightest Ly$\bmath{\alpha}$ emitters at $\bmath{z\sim 5.7}$
\thanks {Based on
    observations obtained at the European Southern Observatory using
    the ESO Very Large Telescope on Cerro Paranal through ESO programs
    275.A-5012, 076.A-0553 and 080.A-0237 and on observations taken
    with the Spitzer Telescope through GO programs 50308 and 60059}}

\author[C. Lidman et al.]
  {C. Lidman,$^1$\thanks{E-mail: clidman@aao.gov.au}
  M. Hayes,$^{2,3}$, D. H. Jones,$^4$ D. Schaerer,$^{2,3}$ E. Westra,$^5$ C. Tapken,$^6$
  \newauthor
   K. Meisenheimer,$^7$ and A. Verhamme$^{8}$\\
  $^{1}$Australian Astronomical Observatory, PO Box 296, Epping NSW 1710, Australia\\
  $^{2}$Observatoire de Gen\`eve, Universit\'e de Gen\`eve, 51 Ch. des Maillettes, 1290 Versoix, Switzerland\\
  $^{3}$CNRS, IRAP, 14 Avenue E. Belin, F-31400 Toulouse, France\\
  $^{4}$School of Physics, Monash University, Clayton, VIC 3800, Australia\\
  $^{5}$Smithsonian Astrophysical Observatory, 60 Garden Street, Cambridge, MA 02138, USA\\
  $^{6}$Astrophysikalisches Institut Potsdam, An der Sternwarte 16, D-14482 Potsdam, Germany\\
  $^{7}$Max-Planck-Institut f\"{u}r Astronomie, K\"{o}nigstuhl 17, D-69117 Heidelberg, Germany\\
  $^{8}$Universit\'{e} de Lyon, Lyon, F-69003, France; \\Universit\'{e} Lyon 1, Observatoire de Lyon, 9 avenue Charles Andr\'{e},Saint-Genis Laval, F-69230, France; \\CNRS, UMR 5574, Centre de Recherche Astrophysique de Lyon, Ecole Normale Sup\'{e}rieure de Lyon, Lyon, F-69007, France
}

\begin{document}

\date{Accepted YYYY Month DD. Received 2011 April 20}

\pagerange{\pageref{firstpage}--\pageref{lastpage}} \pubyear{2002}

\maketitle

\label{firstpage}

\begin{abstract}
  We use deep VLT optical and near-IR spectroscopy and deep
  Spitzer/IRAC imaging to examine the properties of two of the most
  luminous \lya emitters at $z=5.7$.  The continuum red-ward of the
  \lya line is clearly detected in both objects, thus facilitating a
  relatively accurate measurement (10--20\% uncertainties) of the
  observed rest-frame equivalent widths, which are around
  $160\,\mathrm{\AA}$ for both objects. Through detailed
  modelling of the profile of the \lya line with a 3-D Monte-Carlo
  radiative transfer code, we estimate the intrinsic rest-frame
  equivalent width of \lya and find values that are around
  300\,$\mathrm{\AA}$, which is at the upper end of the range allowed
  for very young, moderately metal-poor star-forming
  galaxies. However, the uncertainties are large and values as high as
  700\,$\mathrm{\AA}$ are permitted by the data. Both \lya emitters
  are detected at 3.6\,\mic in deep images taken with
  the {\it Spitzer Space Telescope}. We use these measurements, the measurement of the
  continuum red-ward of \lya and other photometry to constrain the
  spectral energy distributions of these very luminous \lya emitters
  and to compare them to three similar \lya emitters from the literature.
  The contribution from nebular emission is included in our models:
  excluding it results in significantly higher masses.  Four of the
  five \lya emitters have masses of the order of $\sim 10^9$ \msun\ and
  fairly high specific star-formation rates ($\ga$ 10--100
  Gyr$^{-1}$).  While our two \lya emitters appear similar in terms of
  the observed \lya rest-frame equivalent width, they are quite
  distinct from each other in terms of age, mass and star formation
  history.  Evidence for dust is found in all objects, and emission
  from nebular lines often make a dominant contribution to the rest
  frame 3.6\,\mic flux. Rich in emission lines, these objects are
  prime targets for the next generation of extremely large telescopes,
  JWST and ALMA.

\end{abstract}


\section{Introduction}

Characterising the properties of very high redshift star-forming
galaxies is important for a number of studies. For example, it is now
generally accepted that ionising photons from massive stars in these
galaxies led to an important event in the history of the Universe: the
reionisation of hydrogen in the intergalactic medium (IGM)
\citep{Robertson10a}. With the advent of Wide Field Camera 3 (WFC3) on
the {\it Hubble Space Telescope (HST)}, we are making rapid progress
in discovering significant numbers of these galaxies
\citep{Bunker10a,Bouwens10a,McLure10a}; however, progress on
characterising their properties is slower, largely because these
galaxies are extremely faint.

Detailed spectroscopic follow-up of most of these objects will need to
wait for the upcoming generation of space and ground-based facilities,
as the signal-to-noise ratios that are obtainable with current
facilities are generally too low
\citep[e.g.][]{Capak11a,Lehnert10a,Stark10a}. Alternatively, one can
study nearer and brighter objects that are expected to be low-redshift
analogues of these very high redshift sources
\cite[e.g.][]{Izotov09a}, or, as we do in this paper, study the
brightest high-redshift examples.

Very high redshift galaxies (defined here as galaxies with redshifts
greater than 5) are expected, intrinsically at least, to be powerful
emitters in the \lya line \citep{Hayes10a}.  However, \lya is a
resonance line with a large opacity, so even a small amount of neutral
hydrogen modifies the line profile. Just how much and in what way
depends on the intrinsic properties of the line and the distribution
and kinematics of gas and dust in both the interstellar medium (ISM)
attached to the galaxy and the IGM along the line-of-sight.  Many high
redshift galaxies do indeed emit strongly in the \lya line and are
often called \lya emitters or LAEs for short.

Between 2000 and 2002 we undertook a survey to find the brightest LAEs
at $z=5.7$ with the aim of furnishing a sample that could be studied
in some detail with the new generation of telescopes that were
becoming available at that time. This survey, called WFILAS (Wide
Field Imager Lyman Alpha Survey), was conducted with the ESO 2.2m
telescope and its Wide Field Imager (WFI), in conjunction with three
narrowband filters and one intermediate band filter, all located in a
region of low terrestrial background at $\sim 815$\,nm
\citep{Westra06}. The survey covered 0.74 square degrees and furnished
7 LAEs above a flux threshold of $5 \times 10^{-17}$\,erg/s/cm$^2$.

In this paper, we investigate the properties of two of the brightest
LAEs from WFILAS survey using deep VLT optical and near-IR spectra and
Spitzer/IRAC images. They are among the brightest LAEs currently known
at redshift $z=5.7$, and are therefore amenable to detailed study with
current facilities. In Section 2 of the paper, we describe the VLT and
Spitzer observations. In Section 3, we analyse the spectroscopic data,
fitting the profile of the \lya line, measuring the continuum and
searching for evidence of other emission lines. In Sections 4, we
combine the spectra with the IRAC photometry to model the spectral
energy distributions. We also reanalyse three other bright LAEs from
\citet{Lai07a}. In Sections 6, we summarise out main
findings. Throughout the paper, all equivalent widths are reported in
the rest frame, unless explicitly noted otherwise, and we assume a
flat $\Lambda$CDM cosmology with $\Omega_{\Lambda}=0.73$ and $H_0=71$.

\section{Target Selection and Observations}

With luminosities of $ \sim 3 \times 10^{43}\,\rm{erg\,s^{-1}}$,
SGP~8884 and S11~5236 from the WFILAS survey of \citet{Westra06} are
among the most luminous LAEs currently known at $z=5.7$. From earlier
observations \citep{Westra05,Westra06}, the 2$\sigma$ lower limit on
the EW was above $100\,\mathrm{\AA}$ for both objects, which made them
potential Population III galaxies and therefore interesting targets to
search for Population III signatures \citep[cf.][]{Schaerer02a}. They
are sufficiently bright that they can be observed in some detail with
current instrumentation and before the availability of JWST, ALMA and
the next generation of extremely large ground-based large telescopes.
The properties of these LAEs, taken from \citet{Westra05,Westra06},
are summarised in Table.\ref{table:summary}.

Our observations consisted of deep medium-resolution spectroscopy with
FORS2 and SINFONI on the VLT and 3.6 and 5.8\,\mic imaging with IRAC
on the Spitzer Space Telescope.  The FORS2 spectra cover the rest
frame 1180 to 1400$\,\mathrm{\AA}$ region, which includes Lyman alpha and
NV\,$\lambda$1240. The SINFONI spectra cover the rest frame 1640 to
2000 $\,\mathrm{\AA}$ region, which includes several lines, such as the
CIII]\,$\lambda\lambda$1907,1909 doublet, and, for S11~5236 only,
HeII\,$\lambda$1640.  At $z=5.7$, the [OIII]\,$\lambda\lambda$
4959,5007 doublet and H$\beta$ lie within the passband of the IRAC
3.6\,\mic filter.

\begin{table*}
\caption{Properties of SGP~8884 and S11~5236}
\label{table:summary}  
\centering           
\begin{tabular}{l r r c c c}
\hline
\hline
Object & RA & Declination &   EW$^1$     & Flux (imaging)$^2$ & Flux (spectroscopy)$^3$\\
       &    &             &  [$\,\mathrm{\AA}$] & [$10^{-17}$\,erg\,s$^{-1}$\,cm$^{-2}$] & [$10^{-17}$\,erg\,s$^{-1}$\,cm$^{-2}$]\\  
\hline                           
SGP~8884 & 00:45:25.38 & -29:24:02.8 & $>125$ &  $9.5 \pm 1.4$ & $9.05\pm 0.06$ \\
S11~5236 & 14:43:34.98 & -01:44:33.7 & $>100$ &  $7.0 \pm 1.2$ & $4.18\pm 0.04$ \\      
\hline
\multicolumn{6}{l}{Note 1: Initial $2 \sigma$ lower limits from \citet{Westra05,Westra06}}\\
\multicolumn{6}{l}{Note 2: Narrow band fluxes from \citet{Westra06}} \\
\multicolumn{6}{l}{Note 3: Integrated line flux from this paper. Does not include slit losses}\\
\end{tabular}
\end{table*}


\subsection{FORS2 spectroscopy}\label{sec:FORS2spectro}

SGP~8884 and S11~5236 were observed with the multi-object
spectroscopic (MOS) mode of FORS2 \citep{Appenzeller98} on Antu
(VLT-UT1). The MOS mode of FORS2 consists of 19 movable slits that can
be used to select targets over a 7\arcmin\ field. The slit length
varies between 20\arcsec\ and 22\arcsec. All data were taken with the
FORS2 1028z holographic grating and the OG590 order sorting filter,
which has a throughput of almost 80\% in the region surrounding
810\,nm. The slit width was set to 0\farcs8. The details of the FORS2
observations are listed in Table~\ref{table:FORS2}.

One MOS slit was placed on the LAE. Other slits were placed on a
selection of galaxies and stars. The stars were used to monitor the
spectral PSF. Individual exposures were of 1370 seconds duration, and
two of these were taken in a single observing block (OB). Several of
these OBs were executed over the course of several nights to reach the
necessary depth. To facilitate the removal of bright sky lines and to
reduce the impact of cosmic rays and CCD charge traps, the object was
placed in a slightly different part of the slit for each exposure.

\begin{figure*}
  \centering
  \includegraphics[width=16cm]{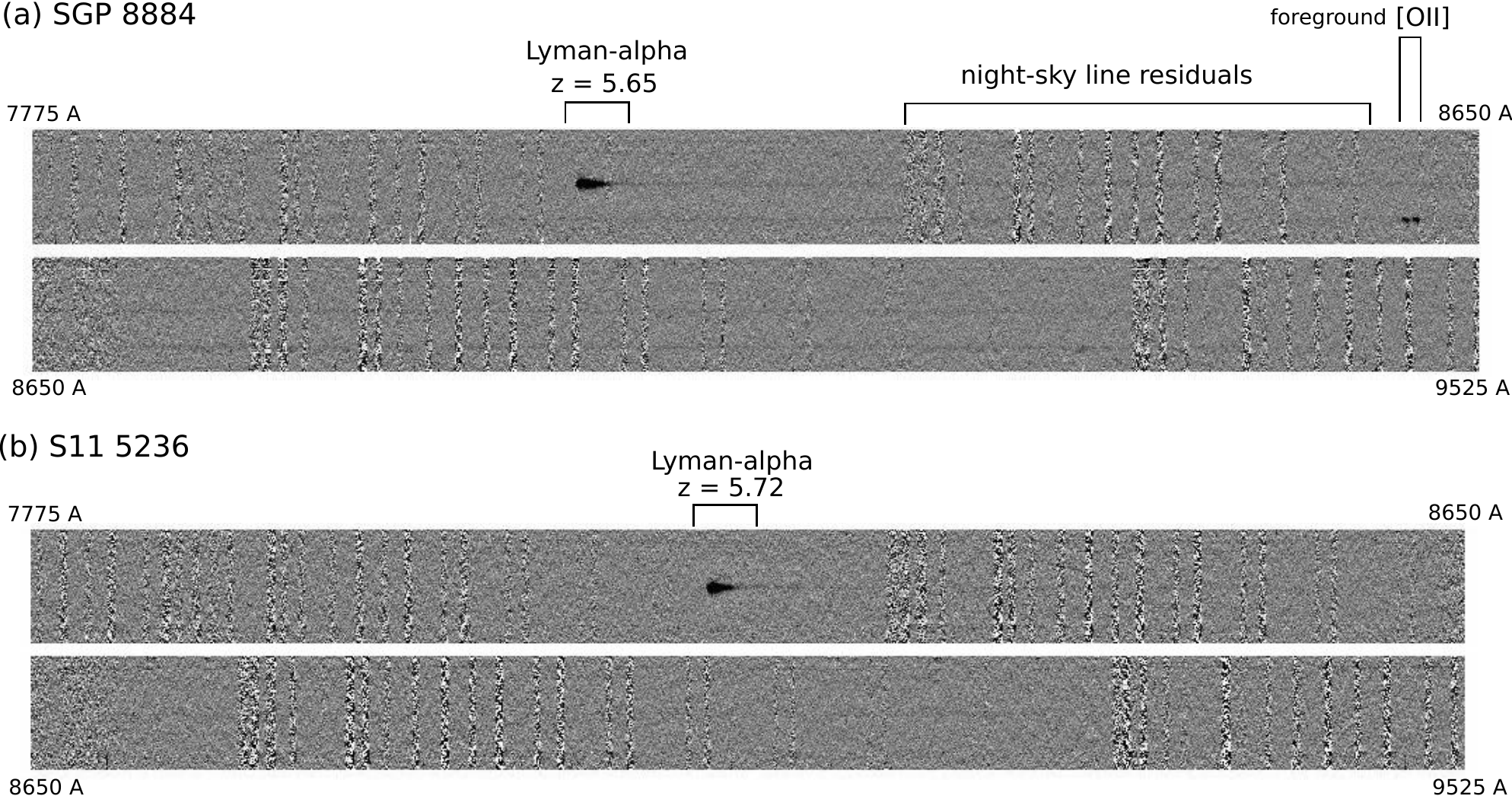}
  \caption{Sky subtracted spectra of SGP~8884 (top two
    panels) and S11~5236 (bottom two panels). The spectra are
    split into two for display purposes. Noisy regions mark the location of bright
    night sky lines. The
    continuum red-ward of the Lyman alpha line is visible in both
    objects, although it is considerably fainter for
    S11~5236. The spectral trace of an [OII] emitting galaxy
    lies about 6\arcsec\ below the spectral trace of SGP~8884. At a
    resolution of 5000, the [OII]\,$\lambda\lambda\,3727$ doublet is
    clearly resolved.}
  \label{figure:2dspectra}
\end{figure*}

The data were reduced in a standard manner using IRAF\,\footnote{IRAF
  is distributed by the National Optical Astronomy Observatories which
  are operated by the Association of Universities for Research in
  Astronomy, Inc., under the cooperative agreement with the National
  Science Foundation} tasks. The bias was removed by fitting a low
order polynomial to the overscan region and pixel-to-pixel variations
were removed with lamp flats. The removal of night sky lines, cosmic
rays and the co-addition of the 2-dimensional spectra were done using
our own software. These sky subtracted and co-added 2-dimensional
spectra are shown in Fig.~\ref{figure:2dspectra}. Although the
continua are considerably fainter than the \lya emission, they are
still visible in the 2-dimensional spectra
(Fig.~\ref{figure:2dspectra}). The 1-dimensional spectra of the LAEs
were then extracted and calibrated in wavelength and flux.

\begin{table}
\caption{FORS2 Observations}
\label{table:FORS2}  
\centering           
\begin{tabular}{l r r}
\hline\hline            
Object & Total exposure time & Median seeing \\
& [s] & [\arcsec] \\
\hline                           
   SGP~8884 & 16440 & 0.65 \\  
   S11~5236 & 43840 & 0.77 \\  
\hline                           
\end{tabular}
\end{table}

The signal-to-noise ratio of the \lya line varies from 100 for
S11~5236 to 150 SGP~8884. The integrated \lya fluxes are within a
factor of two of the fluxes derived from narrow-band imaging
\citep{Westra06}. The details are listed in Table \ref{table:summary}.
The larger difference for S11~5236 can be attributed to the poorer
observing conditions during which the spectra were taken. In the
remainder of this paper, we scale the FORS2 spectra so that the \lya
fluxes computed from spectroscopy match those derived from narrow-band
imaging.

\subsection{SINFONI spectroscopy}

SGP~8884 and S11~5236 were observed with SINFONI on
Yepun (VLT-UT4). SINFONI \citep[]{Eisenhauer03,Bonnet04} is a
near-infrared integral field spectrograph that has an adaptive optics
(AO) module that can be used with natural guide stars or a laser guide
star to improve the spatial resolution.  The observations were done
without AO due to absence of sufficiently close and bright field stars
for AO use.

The targets were observed with the 0\farcs25 objective, which has a
field of view of $8\arcsec \times 8\arcsec$ and 0\farcs125 $\times$
0\farcs25 spaxels, and the $J$ grism, which covers the 1095\,nm to
1350\,nm wavelength range with a resolution of approximately 2000.
Each observing block consisted of four 600 second exposures. Between
exposures, the telescope was offset by a small amount so that the data
could be used to estimate the sky while the object remained within the
SINFONI 8\arcsec\ field of view. The details of the SINFONI observations are listed in
Table~\ref{table:SINFONI}.

The data were reduced following the steps described in \citet{Cuby07}. 
The spectra were extracted with 1\arcsec\ apertures that were placed
at the locations the objects were expected to be. The spectra and the
associated error spectra were calibrated in wavelength and flux. 

The continuum of neither object was detected, even after binning the
extracted 1d spectra. No emission lines were detected either. However,
the SINFONI data allow us to place upper limits on the most likely
emission lines that lie in the region covered by the SINFONI data (see
Sec.~\ref{sec:otherlines}).

\begin{table}
\caption{SINFONI Observations}
\label{table:SINFONI}  
\centering           
\begin{tabular}{@{}lrr}
\hline\hline            
Object & Total exposure time & Median Seeing \\
       & [s]                 & [\arcsec] \\  
\hline                           
SGP~8884 & 24000 &  0.8 \\    
S11~5236 & 9600  &  0.7 \\ 
\hline                           
\medskip
\end{tabular}
\end{table}

\subsection{IRAC imaging and photometry}

SGP~8884 was imaged at 3.6 and 5.8\,\mic with IRAC on the
Spitzer Telescope during cycle 5, a period during which the instruments
were still being cryogenically cooled. To reach the required
sensitivity, the total integration time was set to 36,000 seconds,
which was split into 2 AORs\footnote{Astronomical Observing
  Request}. Each AOR consisted of 180 100-second exposures. Given the
large number of offsets and the characteristics of the target, we
chose the cycling dither pattern with a medium scale factor.

S11~5236 was also imaged with IRAC. However, the observations
were carried out during cycle 6, when Spitzer had run out of cryogens,
so only the 3.6\,\mic data could be taken.  The integration time
and dither pattern were identical to that used for SGP~8884.

Individual integrations were processed using versions S18.7.0 and
S18.18.0 of the basic calibration data pipeline for SGP~8884
and S11~5236, respectively. The processed data were then
combined to an image with a final pixel scale of 0\farcs61, a factor
of two finer than the real pixel scale of the IRAC cameras, using
version S18.3.1 of the {\tt MOPEX} mosaicing software.

\begin{figure}
   \centering
   \includegraphics[width=8.5cm]{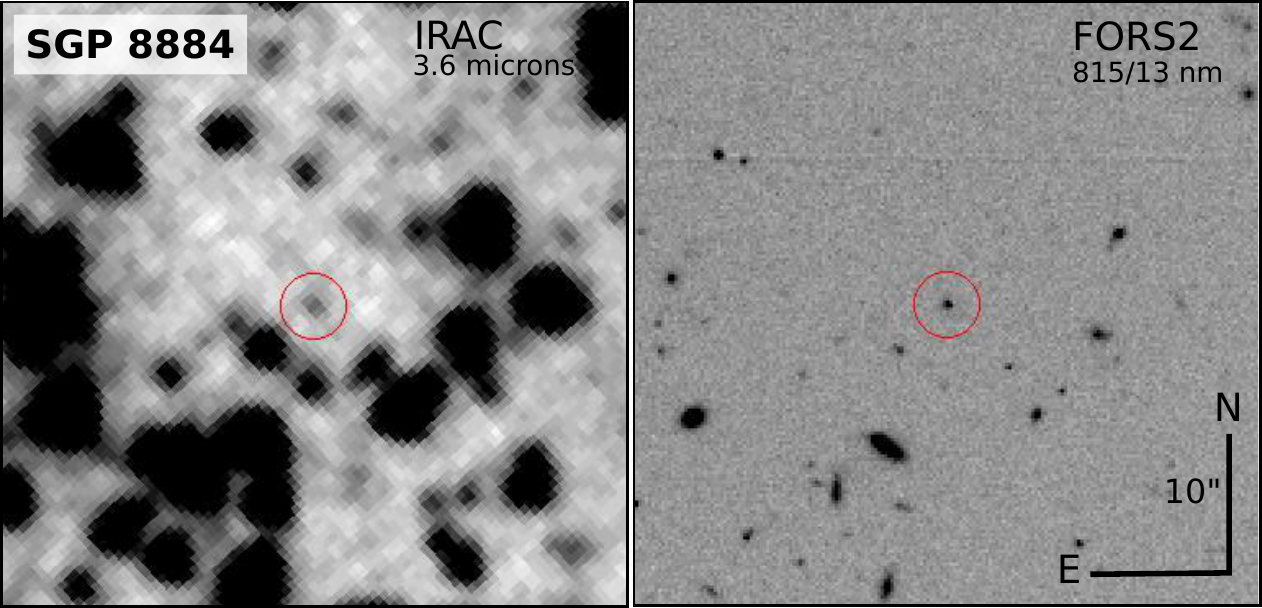}
      \caption{IRAC 3.6\,\mic (left) and FORS2 narrow band (right)
        images of SGP~8884. The images are 45\arcsec across with North
        is up and East is left. The FORS2 image was taken with the
        815-13 narrow band filter and was used to prepare the FORS2
        MOS masks. Most of the flux in the FORS2 image is from \lya. }
      \label{figure:IRAC}
   \end{figure}

For SGP~8884, the IRAC 3.6 and 5.8\,\mic images reach 1\,$\sigma$
point-source sensitivities of 0.03\,$\mu$Jy and 0.32\,$\mu$Jy at the
location of the target.  The 3.6\,\mic image is shown in
Fig.~\ref{figure:IRAC} together with a FORS2 image in the 815-13
narrow band filter that was used to define the MOS mask
(Sec.~\ref{sec:FORS2spectro}). Both images are centred on
SGP~8884. SGP~8884 is undetected in the IRAC 5.8\,\mic image.  

\begin{figure}
   \centering
   \includegraphics[width=8.5cm]{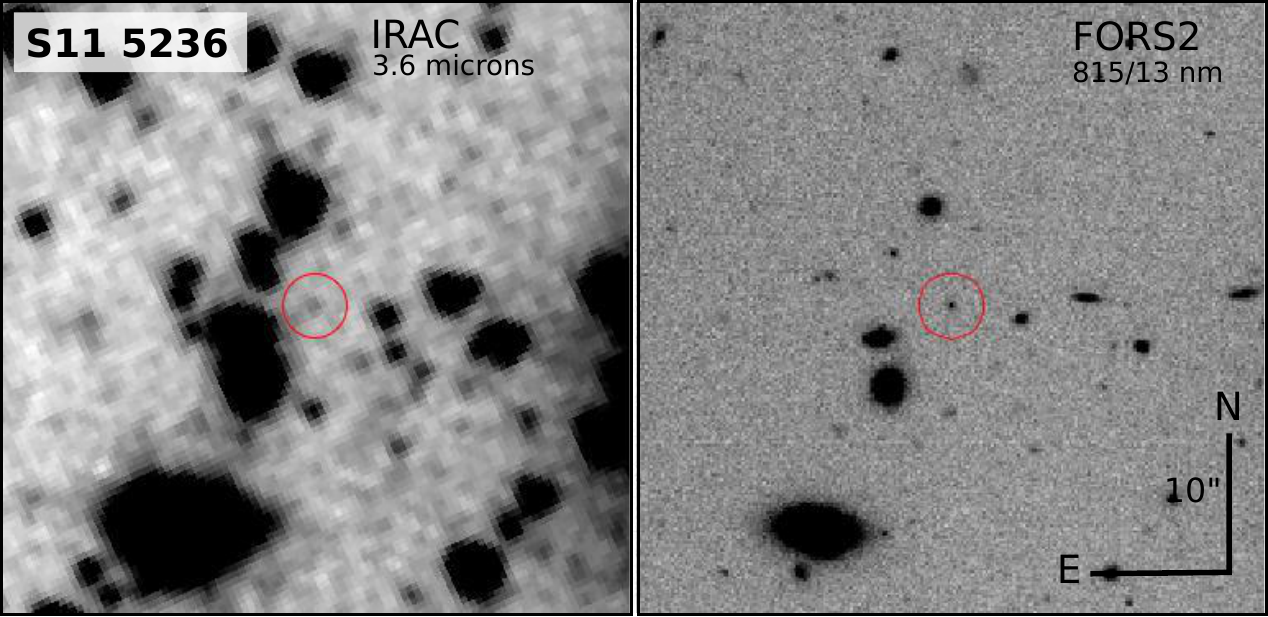}
      \caption{As for Fig.~\ref{figure:IRAC}, but for S11~5236. Most
        of the flux in the FORS2 image is from \lya.}
      \label{figure:IRAC2}
   \end{figure}

For S11~5236, shown in Fig.~\ref{figure:IRAC2}, the 1\,$\sigma$
point-source sensitivity of the 3.6\,\mic image is
0.08\,$\mu$Jy. Since S11~5236 was observed during the warm mission,
there were no 5.8\,\mic data.

Both SGP~8884 and S11~5236 were selected because they were very bright
LAEs and because they lie in relatively uncrowded regions. This is
quite important, as the sensitivity of the IRAC 3.6 micron
observations is well below the limit at which confusion from
overlapping sources becomes a serious issue. The centroids of SGP~8884
in the IRAC [3.6] and FORS2 815-13 images are consistent with each
other, suggesting that there is little contamination from other
objects that happen to land on the line-of-sight. It also means that,
to within the uncertainties in measuring centroids ($\sim0.5$\,kpc at
this redshift), the regions responsible for the \lya emission and the
3.6 flux are coincident. On the other hand, for S11~5236, there is an
offset of approximately 0\farcs5 ($\sim 3$\,kpc) between the two
centroids. This may mean that there is a faint object, not seen in the
FORS2 image, biasing the centroid in the IRAC 3.6\,\mic image or that
the region that is responsible for producing the flux at 3.6\,\mic is
offset from the from the region that produces the \lya emission.

Fluxes were measured in 2\arcsec radius apertures. The centre of the
aperture was determined by the position of LAEs in the FORS2 images
relative to sources that were common to and clearly detected in both
the FORS2 and IRAC images. The fluxes were then corrected to a larger
aperture (radius 4\arcsec) using an aperture correction that was
determined from relatively bright stars. The fluxes are reported in
Table~\ref{table:flux}.

The flux uncertainties are computed
from the uncertainty images produced by {\tt MOPEX}, which includes
only statistical errors. Systematic uncertainties, such as the error
that comes from unresolved sources, are not included. 

\begin{table}
\caption{Fluxes in the observer frame}
\label{table:flux}  
\centering           
\begin{tabular}{lccc}
\hline\hline            
Object      & 0.87\,$\mu$m flux &  3.6\,$\mu$m flux & 5.8\,$\mu$m flux \\
            & ($\mu$Jy)        &   ($\mu$Jy)        &   ($\mu$Jy)      \\
\hline
SGP~8884    &  $0.23 \pm 0.04$ &  $0.52 \pm 0.06$  & $0.32$$^a$\\
S11~5236    &  $0.17 \pm 0.04$ &  $0.48 \pm 0.12$  & ...\\
\hline
$^a 1\sigma$ upper limit
\end{tabular}
\end{table}

\subsection{Morphologies}

Neither SGP~8884 nor S11~5236 are clearly resolved
in the FORS2 images that were used to prepare the masks for
spectroscopy. The image quality was around 0\farcs5 in both
images, so the \lya emitting regions are smaller than 3\,kpc.

However, a field containing S11~5236 was observed with the Advanced
Camera for Surveys instrument (ACS) on HST as part of program 10798
(PI Koopmans) with both the F555W filter, where it is not detected,
and the F814W filter, where it is clearly is detected. The FWHM of a
Gaussian fit to the profile of S11~5236 is 0\farcs2, which is
considerably broader than the PSF and corresponds to physical size
1.2\,kpc at the redshift of S11~5236.

Approximately half of the flux in F814W ACS image of S11~5236 comes
from the \lya line. The rest comes from the continuum red-ward of the
line. 

\section{Analysis of the Spectral Properties}

\subsection{Modelling the continuum and deriving the rest-frame EW of Ly$\bmath{\alpha}$}\label{sec:contFitting}

By inspecting the sky subtracted spectra in
Fig.~\ref{figure:2dspectra}, it is clear that the continuum red-ward
of the Lyman alpha line is detected in both objects.  We fit a power
law with index $\beta$ to the data red-ward of the line and use the
fit to estimate the flux of the continuum at \lya and the observed
rest-frame equivalent width (EW) of the line. The wavelength interval
of the fit is indicated in column 2 of
Table~\ref{table:powerelawfits}. A power law provides an adequate fit
to the data, although given the uncertainties, the error in $\beta$ is
substantial and allows one to constrain neither the age of the stellar
population responsible for the UV emission nor the amount of
reddening. Excluding the SINFONI data from the fit does not change the
value of $\beta$ significantly.

Blue-ward of \lya, the continuum is clearly suppressed relative to the
continuum red-ward of the line. In fact, we do not detect a significant
signal in either object. We fit the average level of the continuum on the
blue side of the line using all the data within the range indicated in
column 6 of Table~\ref{table:powerelawfits}. The mean values for
SGP~8884 and S11~5236 are consistent with zero within 2--$\sigma$.

The ratio between the measured continuum between \lya and \lyb and the
one extrapolated from the continuum red-ward of \lya is often used to
measure the fraction of neutral hydrogen in the IGM. The ratio has been well
measured for QSOs \citep{Songaila04a}, and at $z=5.7$ is around a few
percent, which corresponds to a neutral fraction of $\sim 10^{-4}$,
with some scatter between objects \citep{Fan06a}.

The upper limits for our two LAEs are a factor of five less constraining. If we
extrapolate the continuum red-ward of \lya to 1185$\,\mathrm{\AA}$
(the upper end of the wavelength range used in \citet{Songaila04a}),
we find a flux ratio of $0.23\pm0.11$ for SGP~8884 and $0.14\pm0.11$
for S11~5236. While this indicates that the continuum blue-ward of the line
is affected by neutral hydrogen in the IGM, it does not show that the line
itself is affected. We will return to this issue later in this section.

\begin{table*}
\caption{Quanitites derived from to the fits to the continuum red-ward and blue-ward of \lya.}
\label{table:powerelawfits}  
\centering           
\begin{tabular}{@{\extracolsep{-0.1cm}}lcccc@{\extracolsep{1.5cm}}c@{\extracolsep{0.1cm}}c}
\hline\hline            
Object & Fit interval    & $\beta$ &  Continuum flux at \lya                   & EW                 & Fit interval & Continuum flux below \lya \\
       & [$\mathrm{\AA}$]&         &  [1e-20 ergs/s/cm$^2$/$\,\mathrm{\AA}$]   & [$\,\mathrm{\AA}$] & [$\mathrm{\AA}$] & [1e-20 ergs/s/cm$^2$/$\,\mathrm{\AA}$] \\
(1)&(2)&(3)&(4)&(5)&(6)&(7)\\
\hline                           
SGP~8884 & 1222--1797 & $-1.7 \pm 1.5$ & $8.2\pm 1.0$                     & $166 \pm 20 $  & 1169-1214    & $1.9\pm 0.9$\\
S11~5236 & 1243--1778 & $-3.7 \pm 2.9$ & $3.9\pm 0.8$                     & $160 \pm 36 $  & 1147-1214    & $0.6\pm 0.5$ \\
\hline                           
\end{tabular}
\end{table*}

\subsection{Ly$\bmath{\alpha}$ fitting \label{linefitting} }

The spectra of both LAEs, centred on \lya, are shown in
Fig.~\ref{fig:simplefit}. Both lines are clearly asymmetric, with
steep increases on the blue side of the lines and long red tails.
Very likely, the asymmetry is indicative of an expanding ISM
\citep{Verhamme06}. A relatively weak shoulder on the red side \lya
line of S11~5236, tentatively identified in \citet{Westra05}, is now
clearly visible. We fit the line with a variety of models, first
trying analytic models that have been used in the past and then more
sophisticated 3-D Monte Carlo radiative transfer models.

\subsubsection{Analytic models \label{sec:analytic}}

\begin{figure*}
   \centering
   \includegraphics[width=8.5cm]{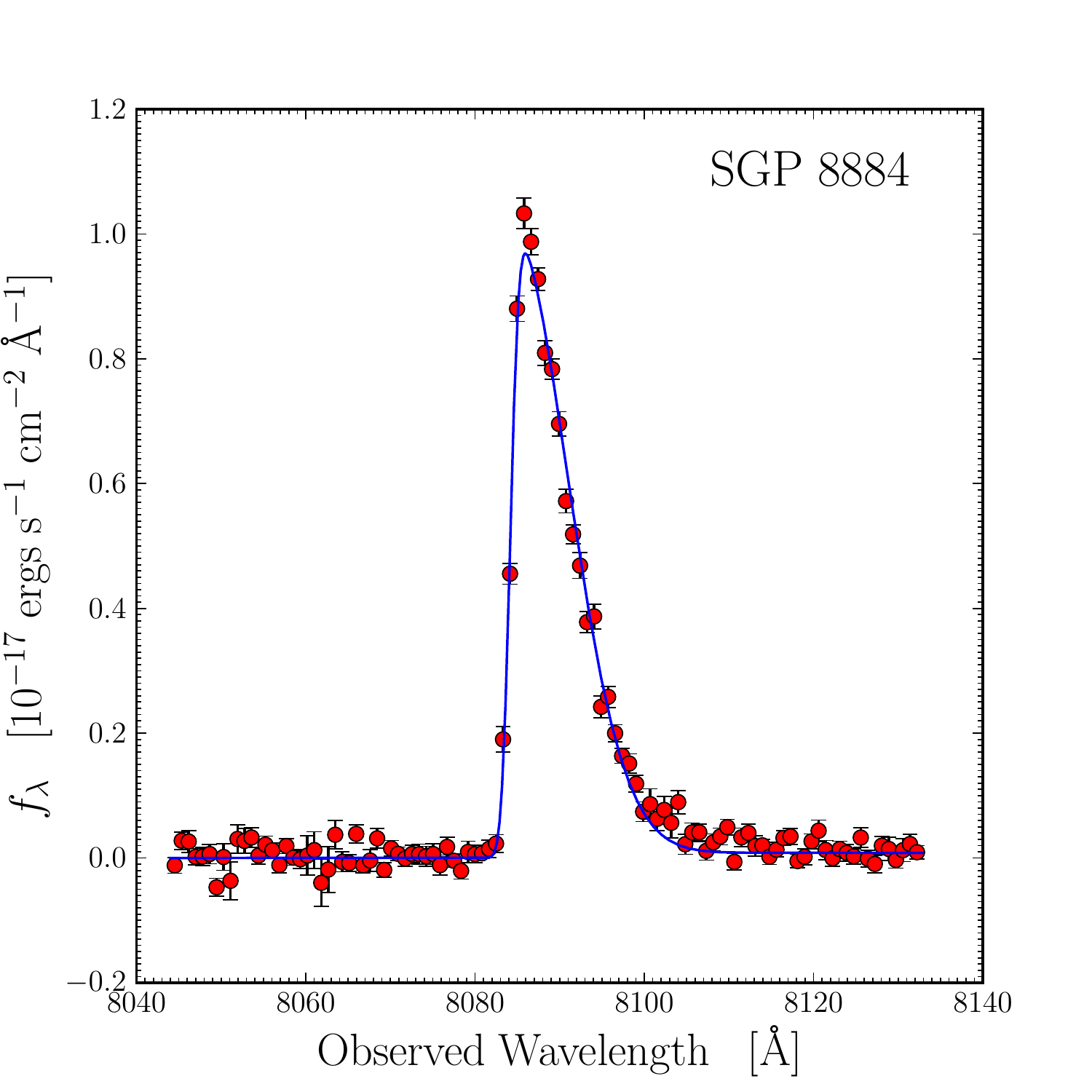}
   \includegraphics[width=8.5cm]{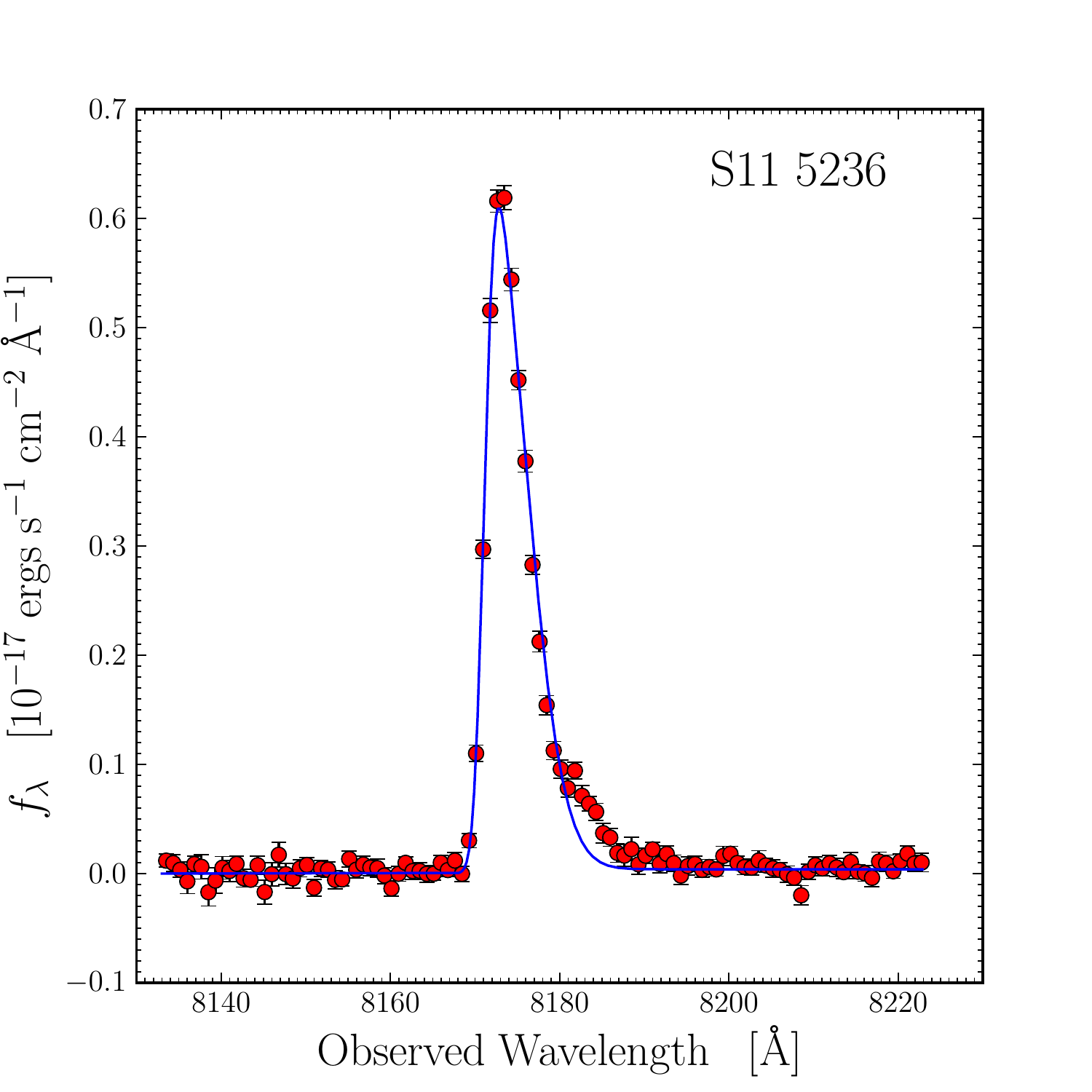}
      \caption{
					{
					Model fits to SGP~8884 (left) and S11~5236 (right). 
                                        The analytic model used in the fits is shown as the blue solid line.
					VLT/FORS2 data are plotted as red points.
				}
			}
      \label{fig:simplefit}
   \end{figure*}

Following \citet{Dawson02a,Hu04} and \citet{Westra05,Westra06}, we fit
the profile of the \lya line with a truncated Gaussian that is
convolved with the instrument resolution. There are four parameters in
the fit: the FWHM and centre of the truncated Gaussian, the instrument
resolution, which is itself modelled as a Gaussian, and the overall
normalisation.  Since the average seeing during the observations was
narrower than the width of the slit, the instrument resolution is
mostly defined by the seeing. The continuum blue-ward of the line
centre is set to zero. Red-ward of the line, the continuum is set to
the value determined in Sec.~\ref{sec:contFitting}. The fits are shown
in Fig.~\ref{fig:simplefit}, and the best fit values for the
parameters of the model are shown in Table~\ref{table:line}. The fits
are constrained using the entire wavelength shown in
Fig.~\ref{fig:simplefit}.

On the blue side of the line, the models fit the data very well. The
instrument resolution, which is determined during the fit, is
  around ($R\sim 5000$) and agrees with the reported seeing. The blue
side of the line is consistent with a vertical cutoff. \citet{Ouchi10}
reported the detection of a ``knee'' on the blue side of the profile
in LAEs at $z=5.7$ and $z=6.5$. Our data have considerably better
signal-to-noise and better resolution ($\sim 1.6\,\mathrm{\AA}$).
There is no evidence for such a feature in the two LAEs that we
observed.

On the red side, however, the models are a poor fit and fail to
reproduce the level of detail seen in the data.  For S11~5236, the
shoulder at $\sim 8183\,\mathrm{\AA}$ is, not surprisingly, completely
missed. The fit for SGP~8884 is better; however, there also appears to
be excess flux in the red wing of the line around $\sim
8105\,\mathrm{\AA}$.  Such features are characteristic of an expanding
dusty ISM \cite[cf.][]{Verhamme08}. We now move to using a 3-D Monte
Carlo radiation transfer code to model escape of \lya photons from the
LAEs and to see if we can reproduce the red wing better.

\subsubsection{3-D \lya radiation transfer\label{sec:3d}}

To model the \lya profiles, we use an updated version of the 3-D \lya
radiation transfer code {\tt MCLya} of \citet{Verhamme06} described in
\citet{Schaerer11}. The code has been used successfully in
\citet{Verhamme08,Schaerer08,DZ10a,Vanzella10a} to model Lyman break
galaxies (LBGs) and LAEs. We follow \citet{Verhamme08} and fit a
simple model to the data. The model consists of a line and continuum
emitting region that is surrounded by an expanding, dusty shell of
neutral hydrogen. The dust is assumed to be uniformly distributed
within the shell. There are seven parameters in the model.  In
practice we use an automated fitting tool, which synthesises \lya
profiles from the extensive library of more than 6000 radiative
transfer models of \citet{Schaerer11} covering a wide range of
parameter space.  The continuum at \lya is fixed to the value listed
in Table~\ref{table:powerelawfits} as these have been determined over
a larger wavelength region than the regions used to fit \lya and is
therefore more accurate.

At $z=5.7$, it is possible that resonant scattering of \lya photons by
neutral hydrogen external to the galaxy -- either in the form of
nearby \lya forest systems or remnant neutral gas in the IGM -- may
affect the \lya profile. Stochastic as this is, it provides a Monte
Carlo problem all of its own, and one that we do not have the
possibility to address on a case-by-case basis. Instead, we adopt two
simulations per galaxy, which are designed to bracket a broad range of
possibilities: we firstly assume that the IGM does not affect the
line\footnote {This does not exclude the possibility that the IGM
  starts to suppress the continuum at some point that is blue-ward of
  the line.}, and secondly assume that the IGM absorbs all the flux
blue-ward of the line centre.  In this second approach the
intervention occurs after radiation transport (to maintain the effects
of the ISM) but before convolution with the instrumental profile, and
thus is somewhat comparable to the method used in
Sec.~\ref{sec:analytic}, but with the shape of the red side determined
by simulation.  We present the results from the first simulation
first. The best fit parameters and the resulting escape fractions of
\lya photons and the quality of the fits as characterised by the
reduced $\chi^2$ are shown in Table~\ref{table:line}. The escape
fraction is defined as the fraction of \lya photons that escape the
galaxy. It does not include the fraction that are scattered by the
IGM.

\begin{table*}
  \caption{Line fitting results. Columns 3 through to 9 are obtained from the fits and are, 
    respectively, the redshift ($z$), the expansion velocity 
    of the H{\sc I} shell ($V_\mathrm{exp}$), the column density of neutral hydrogen in the shell ($N(H)$), the 
    amount of dust absorption ($\tau$), a Doppler parameter describing the random motions of the neutral gas ($b$), 
    the intrinsic FWHM of the \lya line (FWHM$_{0}$), and the intrinsic equivalent width of the \lya line (EW$_0$).
    The escape fraction $f_\mathrm{esc}$, is derived from the fitted parameters.}
\label{table:line}  
\centering           
\begin{tabular}{ccccccccccc}
\hline\hline            
Object   & Fit Type & $z$   &$V_\mathrm{exp}$       &$\log N(\mathrm{HI})$              &$\tau$             &$b$                & FWHM$_{0}$        & EW$_0$           &$f_\mathrm{esc}$         & $\chi^2_\nu$ \\
         & &        &[$\mathrm{km\,s^{-1}}$]&  [$\mathrm{cm^{-2}}$]                  &                   &[$\mathrm{km\,s^{-1}}$]&$\,\mathrm{\AA}$             &$\mathrm{\AA}$ &  &  \\
(1) &(2)&(3)&(4)&(5)&(6)&(7)&(8)&(9)&(10)&(11)\\
\hline                           
SGP~8884    & Analytic. &$5.6527$ & ... & ...  & ... & ...& ... & ... & ...    & 2.1 \\
SGP~8884    & {\tt MCLya} &$5.6517$ & 250 & 19.3 & 3.0 & 80 & 300 & 350 & 0.0232 & 2.71 \\
SGP~8884+IGM& {\tt MCLya} &$5.6517$ & 50  & 16.0 & 2.0 & 40 & 600 & 350 & 0.104 & 1.68\\
\hline
S11~5236    & Analytic &$5.7242$ & ... & ...  & ... & ...& ... & ... & ...    & 2.9 \\
S11~5236    & {\tt MCLya} &$5.7245$ & 300 & 18.5 & 1.0 & 80 & 200 & 250 & 0.300 & 2.03 \\
S11~5236+IGM& {\tt MCLya} &$5.7237$ & 400 & 18.0 & 0.2 & 80 & 300 & 350 & 0.819 & 1.01 \\
\hline                           
\end{tabular}
\end{table*}

The redshifts are well constrained and we use these values throughout
the remainder of the paper. Column densities of neutral hydrogen are
determined to be small, just $0.3 \times 10^{19}$~cm$^{-2}$ and $2
\times 10^{19}$~cm$^{-2}$ for S11~5236 and SGP~8884, respectively.
These values are unconstrained at the low $N(\mathrm{HI})$ end, but at
the high end are strongly constrained, with statistical upper
error-bars of 0.4 dex.  The characteristic features that allow more
accurate determination of effective dust optical depth, $\tau$, do not
appear in either of these spectra, so the amount of absorption is
poorly constrained, although reasonable lower-limits are obtained for
both galaxies: we determine lower limits of $\tau=0.5$ for SGP~8884
and $0.7$ for S11~5236 at the 68.3\% confidence limit, although the
upper limit is unconstrained by our grid, even at the 68.3\%
confidence limit. For both galaxies the full range is permitted at the
99.7\% confidence limit.
		
The \lya\ escape fractions are found to be small for SGP~8884 and
moderate for S11~5236, taking values of 2.3 and 30\%,
respectively. For SGP~8884 this value is not tightly constrained and
$f_\mathrm{esc}=50\%$ (100\%) are permitted at the 68.3\% (99.7\%)
confidence limit.  However for S11~5236 $f_\mathrm{esc}$ is
constrained to $<50\%$ at the 99.7\% confidence level.

The intrinsic rest-frame equivalent widths of the \lya line is
350$\,\mathrm{\AA}$ for SGP~8884 and 250$\,\mathrm{\AA}$ for S11~5236,
with values as high as 400 and 500$\,\mathrm{\AA}$ permitted for
S11~5236 and SGP~8884, respectively (68.3\% confidence). The lower
limits just reproduce the measured EWs given in
Table~\ref{table:powerelawfits}.

\begin{figure*}
   \centering
   \includegraphics[width=8.5cm]{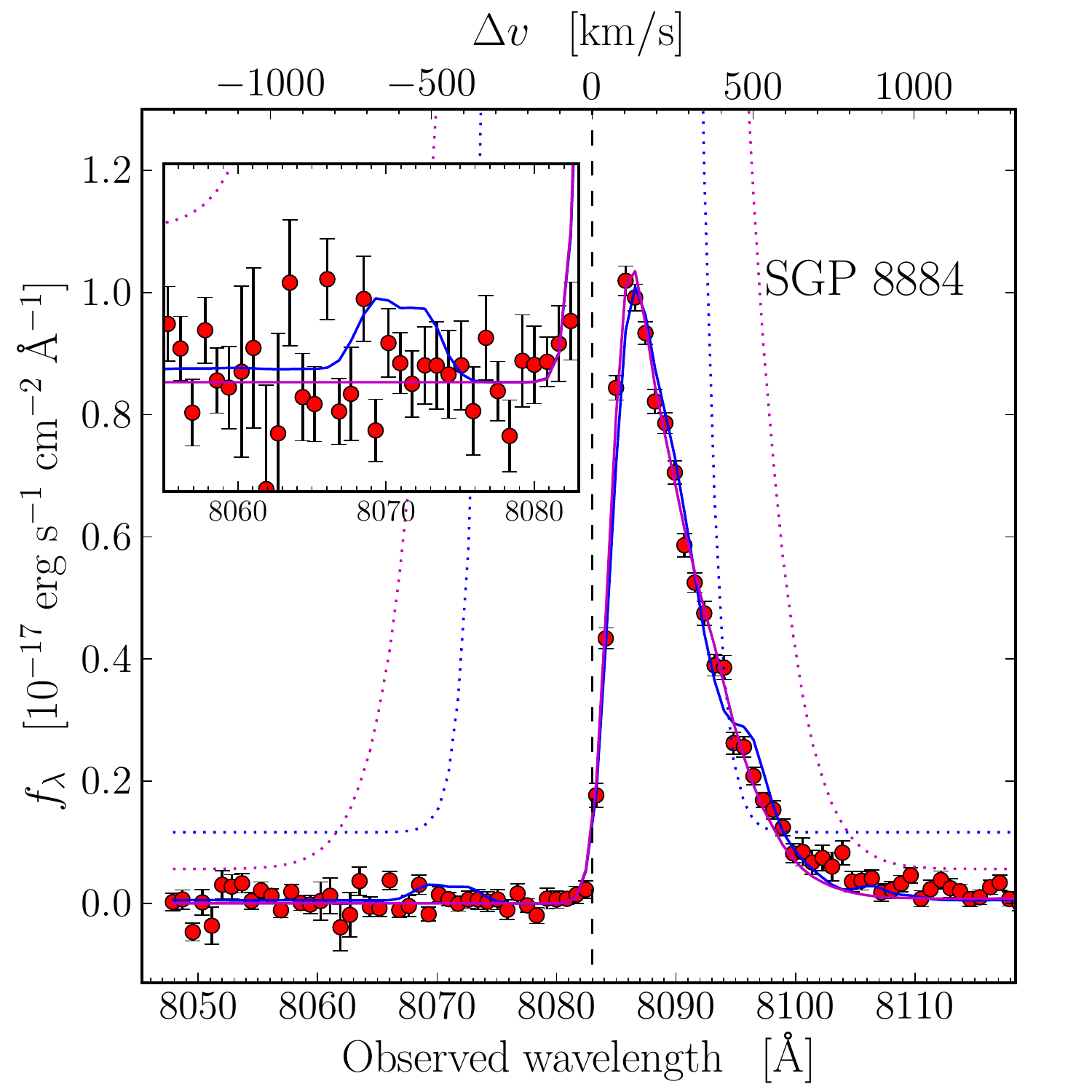}
   \includegraphics[width=8.5cm]{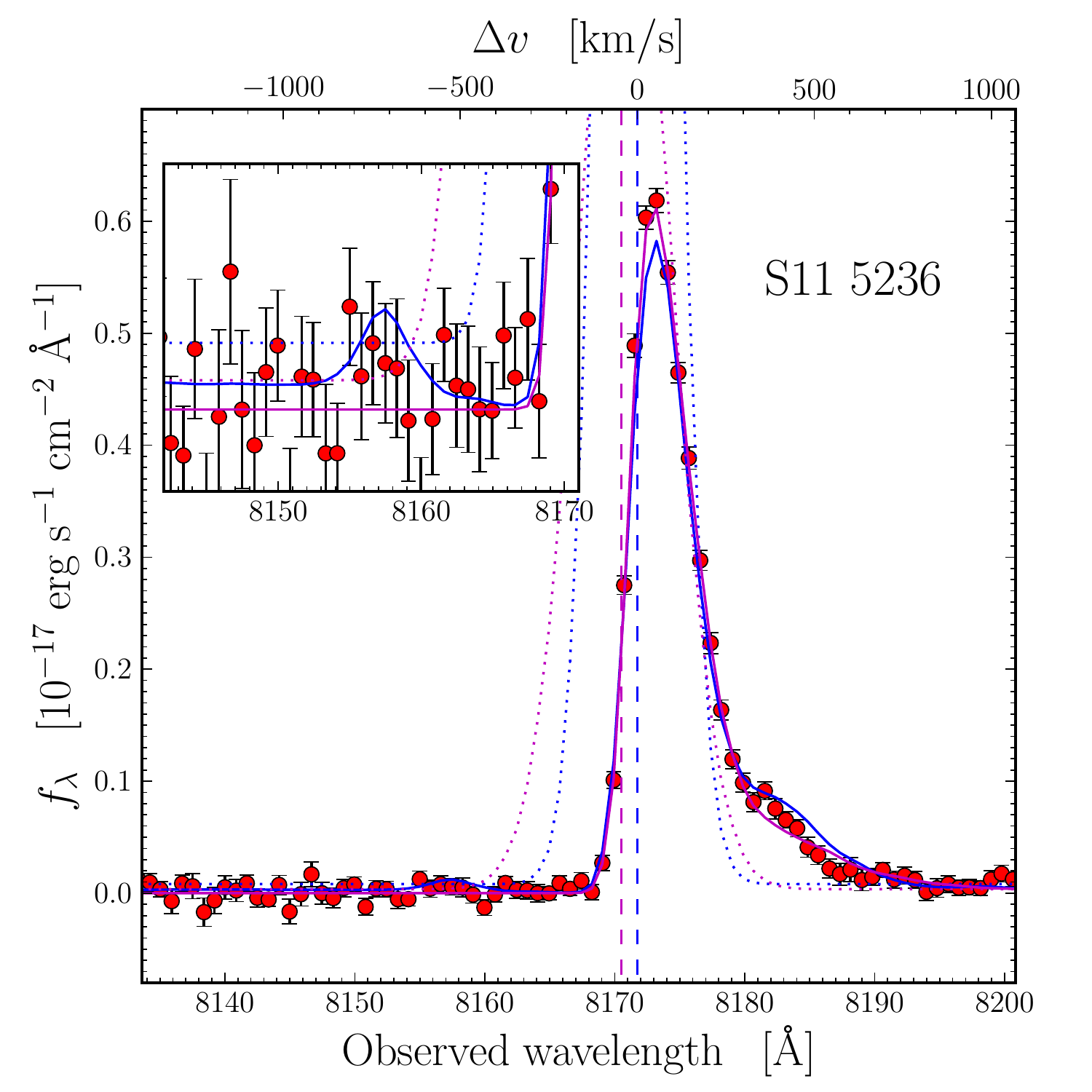}
      \caption{
					Model fits to SGP~8884 (left) and S11~5236 (right). The model 
					in which the IGM attenuation is ignored is shown by the blue 
					unbroken line, while the magenta line shows the model for which
					the transmission through the IGM was set to zero blue-ward of 
                                        the systemic redshift. 
					The intrinsic spectra (i.e.~before radiation transport 
                                        and attenuation) is shown 
					by the dotted line in both cases. 
					VLT/FORS2 data are plotted as red points. The vertical dashed lines mark the
                                        line centre of \lya for the four cases. For SGP~8884, the line centre of \lya 
                                        overlaps in the two cases considered. The velocity scale in
                                        the right hand figure is only valid for the case where IGM attenuation is 
                                        ignored. The insets illustrate magnified views of the continuum blue-ward of \lya,
                                        approximately within 1000\,\kms of line centre.
			}
      \label{figure:line}
   \end{figure*}

The radiation transfer code (the blue curve in Fig.~\ref{figure:line})
provides a better fit to the the main features of the \lya line in
S11~5236 than the analytic model used in Sec.~\ref{sec:analytic}.  In
particular, the red shoulder that is seen in S11~5236 is reproduced by
the model.  Conversely, for SGP~8884, the radiation transfer code
provides a poorer fit before the IGM is considered.

For both LAEs, the simulations suggest that at wavelengths short-ward
of the \lya\ line ($\sim$ -700 to -800 \kms) the effects of
radiation transport on continuum photons become negligible, and thus
(modulo the dust extinction law) should behave no differently from
photons on the red side of \lya\ (see Fig.\ \ref{figure:line}).  At
this scale, i.e.\ close to the \lya\ line the model is compatible with
the observations.

The second of our simulations assumes that all the transmitted flux
blue-ward of the systemic velocity is absorbed by the IGM.  Obviously
this removes all flux, including the small bumps, on the blue side of
the line for both galaxies (magenta line cf.\ blue line).  However,
even with the quality of the present spectra it is not possible to
distinguish from the observations blue-ward of \lya\ whether or not
the IGM affects directly the \lya\ line. Both models fit the blue side
of the \lya line equally well. In principle, the IGM may also affect
the main red peak by giving the impression that the peak itself has
shifted red-ward. This effect, anticipated by, for example,
\citet{Haiman02,Santos04} and \citet{Dijkstra10} and noted in
\citet{Verhamme08}, appears to be largely mitigated by the fact that
the line is already shifted to the red from the multiple scatterings
that occur in the ISM.

Quantitatively the quality of the overall line fit is improved for
both objects by the addition of the IGM, bringing $\chi^2$ per
degree-of-freedom down to 1.7 and 1.0 for SGP~8884\footnote{ For
  SGP~8884, there is some evidence from the scatter of the data in the
  continuum regions that the errors have been underestimated by
  20--30\%.}  and S11~5326 respectively and more in line with the
values that we are looking for in this kind of analysis.  The
treatment of the IGM has further noticeable effects on the derived
parameters that are worthy of discussion. Firstly it substantially
reduces the dust optical depth of the ISM in both galaxies, although
the formal level of constraint is no better than it was in the case
where the IGM was ignored. In the ISM only model, the only way to
remove emitted photons from the system is to absorb them by
dust. Adding the IGM introduces an additional sink of photons. Thus,
when we introduce the step-function IGM, we see higher escape
fractions and consequently lower measures of $\tau$. Indeed the escape
fractions have increased substantially, from 2.3 to 10\% in SGP~8884
and from 30 to 80\% in S11~5326. Interestingly, the best fit intrinsic
rest frame EWs are largely unchanged. The 68.3\% confidence interval
covers a slightly larger range, from 150 to 700$\,\mathrm{\AA}$.

The column density of neutral hydrogen, $N(\mathrm{HI})$, is
substantially reduced for both LAEs; however, the 68.3\% confidence
interval brackets the best fit values of the first simulation.  In the
case of SGP~8884 the intrinsic FWHM of \lya\ doubles to 600 km/s;
however, the uncertainty in this parameter is large. The 68.3\%
confidence interval extends down to 250 km/s. Such broad lines have
been reported in other LAEs
\citep[e.g.][]{Verhamme08,Vanzella10a}. They are more commonly
observed in AGN; however, we see no evidence for AGN activity in
either SGP~8884 or S11~5236. The NV\,$\lambda1240$ line, which is
commonly seen in AGN, was not detected in either LAE. Upper limits on
NV are listed in Table~\ref{table:limits}.

\subsubsection{Discussion}

Neither of the IGM models used in this paper are likely to represent
the average state of the IGM along the line-of-sight to LAEs at
$z=5.7$.  The first model assumes no attenuation of \lya from the IGM. The
probability that this occurs for both objects is quite small
\citep{Laursen11a}. It is worth noting; however, that both LAEs were
selected as being the brightest LAEs over a relatively large area of
sky, so there will be a bias in selecting LAEs that have more
transparent lines of sight than the average. The second model assumes
that all photons blue-ward of the systemic redshift are scattered by
the IGM. This step function neglects the damping wing of \lya, so the
attenuation may extend to longer wavelengths depending on the
kinematics of the IGM with respect to the source and the
ionization state of the circum-galactic medium
\citep[see, for example,][]{Haiman02,Santos04,Laursen11a}.

A more precise treatment of the IGM \citep[as in, for
  example,][]{Laursen11a} is not warranted with the current data. One
clear effect from adopting more accurate IGM models is that some of
the fitted parameters would almost certainly change. For example, if
the IGM were to absorb flux red-ward of the systemic velocity
\citep[as it does in the models of ][]{Laursen11a}, then the best fit
redshift would likely decrease, as it did for S11~5236. Hence, an
direct measure of the redshift from other spectral features, such as
nebular lines in the near-IR, would be useful, as this would reduce
the number of fitted parameters. Unfortunately, our attempts at
detecting such lines with SINFONI proved unsuccessful
(Sec.~\ref{sec:otherlines}).

Similarly, a measurement of the expansion velocity of the shell from
the velocity offset of low-ionisation interstellar lines would reduce
the number of parameters further. Our data were not deep enough to see
these lines. 

\subsection{Other emission lines}\label{sec:otherlines}

In addition to \lya, the FORS2 and SINFONI spectra cover a number of
lines commonly observed in AGNs and LBGs. These include, NV\,$\lambda
1240$, CIII]\,$\lambda\lambda 1907,1909$ and, for S11~5236 only,
  HeII\,$\lambda 1640$. The spectra do not cover the
  NIV]\,$\lambda\lambda 1483,1487$ doublet, which has been seen in
    some star-forming galaxies \citep{Vanzella10a}. Unlike \lya, these
    lines are relatively unaffected by the ISM and IGM, and can be
    used to derive redshifts.

Apart from \lya, we do not detect any line emission in either SGP~8884
or S11~5236.  To derive upper limits to the fluxes and to the line
ratios with respect to the observed \lya flux, we compute the noise in
$9\,\mathrm{\AA}$ bins centred at the redshifted wavelengths of these
lines using the redshifts derived from the fitting of the \lya line in
Section \ref{linefitting}. In Table~\ref{table:limits}, we quote
3-$\sigma$ upper limits. The sensitivity limits vary strongly between
lines because some lines overlap regions of bright telluric emission
from OH and ${\mathrm O}_{\mathrm 2}$, while others are in relatively
clear regions.

In addition to listing CIII] and HeII, we provide
3-$\sigma$ upper limits for OIII] and NV. CIII], OIII] and HeII were
detected in the Lynx arc \citep{Fosbury03} and in Q2323-BX418 \citep{Erb10a}, 
a young,
low metallicity, unreddened galaxy at z=2.3, while NV is
commonly found in AGNs \citep{vandenBerk01a}. While both the Lynx arc and
Q2323-BX418 are LAEs, they appear to be quite different to the average LAE. In addition
to being rich in emission lines, the ionisation parameter for both objects 
is quite high, $U=-1$.

In Table~\ref{table:limits}, we list the flux ratios between \lya
and HeII, CIII] and OIII] in the Lynx arc. If the properties of
SGP~8884 and S11~5236 were similar to the Lynx arc, then both CIII] and OIII]
would have been clearly detected in SGP~8884 and marginally detected in S11~5236. In this respect, both SGP~8884 and S11~5236
are different to the Lynx arc and more similar to the average LAE.

In Sec.~\ref{sec:SED}, we fit SED models that include nebular emission
lines to the broadband photometry. We can check to see if the upper
limits listed for the lines in Table~\ref{table:limits} are consistent
with the strength of these lines in best fit SEDs. For both S11~5236,
where the lines in the SED model are very weak, and for SGP~8884,
where the lines are stronger, this is indeed the case.

\begin{table}
\caption{Line ratios of other lines with respect to the observed \lya flux. Upper limits are 3-$\sigma$.}
\label{table:limits}
\centering           
\begin{tabular}{rcrrr}
\hline\hline            
Line& $\lambda_\mathrm{rest}$ & S11~5236 & SGP~8884 & Lynx\\
\hline
 NV    & 1240 & $ < 0.043$ & $ < 0.010 $ & $ < 0.084$ \\
 HeII  & 1640 & $ < 0.198$ & ...         & 0.033 \\
 OIII] & 1661 & $ < 0.191$ & $ < 0.109 $ & 0.058\\
 OIII] & 1666 & $ < 0.133$ & $ < 0.042 $ & 0.121\\
 CIII] & 1907 & $ < 0.148$ & $ < 0.085 $ & 0.113\\
 CIII] & 1909 & $ < 0.124$ & $ < 0.106 $ & 0.077\\
\hline
\end{tabular}
\end{table}


\section{SED fitting}\label{sec:SED}

To gain further insight into the properties of our objects and other LAEs,
we have carried out SED fits using the {\em Hyperz} code
\citep{hyperz} modified by \citet{SdB09} to take into account
the effects of nebular emission (lines and continua) on broad
band photometry. These effects are expected to be largest for
objects with emission lines (as the case for the LAE discussed 
here) at high redshift, since the contribution of emission lines
to photometric bands increases with $(1+z)$.
The code has already been successfully applied to several samples
\citep[see e.g.][]{SdB10,DZ11}, and the method has also
been implemented by other groups and applied to different samples,
in particular to LAEs at $z \sim$ 2--3 and at 6 \citep{Ono10,Acquaviva11}.

For the SED fits we use the spectral templates of \citet{BC03}
to which we add nebular emission as described in \citet{SdB09}.
We consider three metallicities between solar and 1/200 \zsun,
exponentially decreasing star-formation histories ($\propto \exp^{-t/\tau}$)
with $\tau$ varying between 10\,Myr and $\infty$ (constant star-formation),
and variable extinction described by the Calzetti law.
We have also considered models with a fixed, constant tar-formation rate
(SFR).
Non-detections are included in the minimisation by setting the observed 
flux to zero and taking the $1 \sigma$ upper limit as the error.
The SED fits gives the resulting fit parameters, age, stellar extinction
\ebv, stellar mass \mstar, SFR.
The probability distribution function of these parameters (and derived
uncertainties) are determined by Monte Carlo simulations,
generating a large number (typically 1000 or more) of realisations
of the SED. The stellar masses and SFRs quoted here are based on 
a Salpeter IMF from 0.1 to 150 \msun. For comparison with a more
realistic IMF including the observed turn-over at low stellar masses
these values must be corrected downward (e.g.\ by a factor 1.8
for a Chabrier-type IMF). 

\subsection{SED modelling of SGP 8884 and S11 5236}

\begin{figure}
   \centering
   \includegraphics[width=9cm]{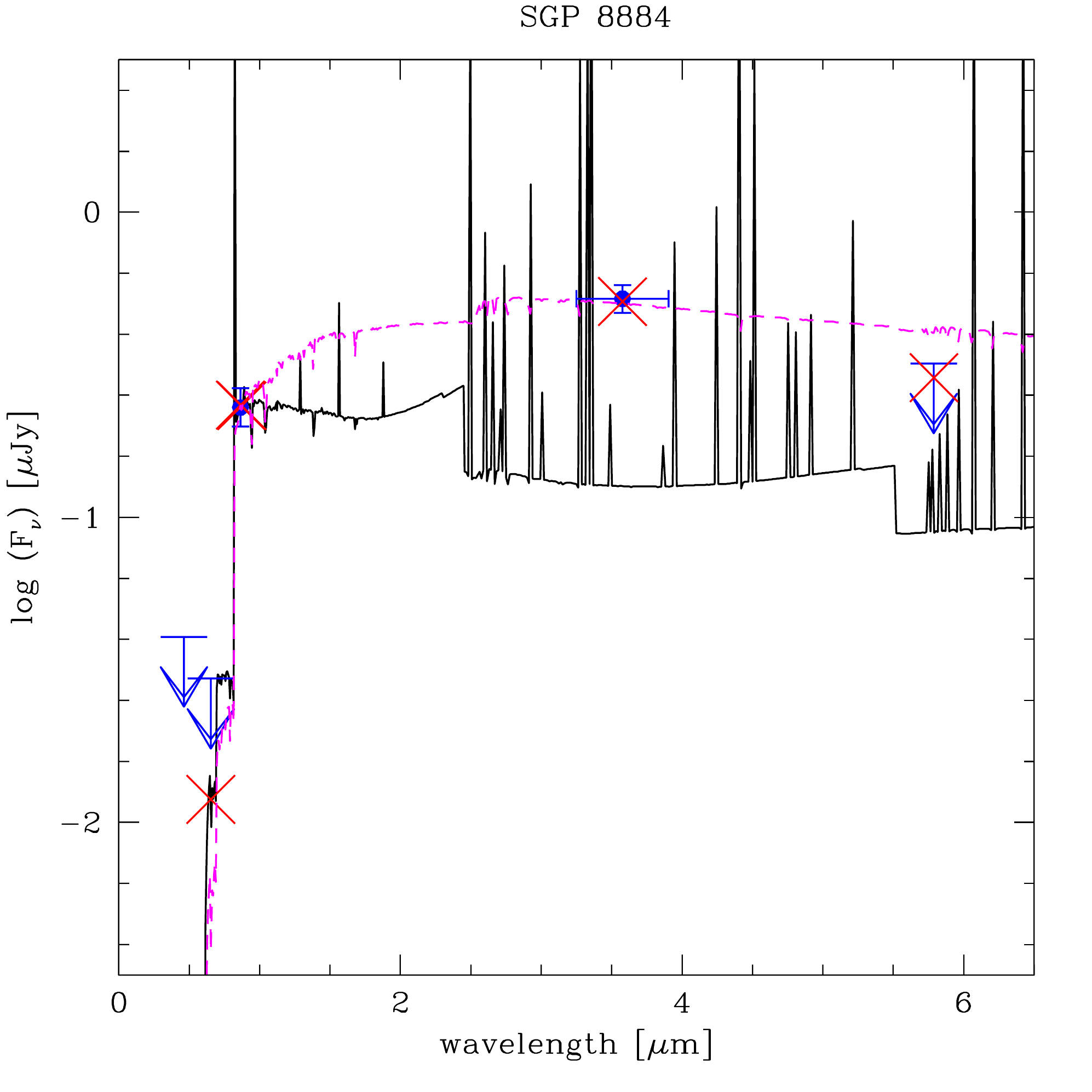}
 
      \caption{Observed (blue symbols) and best-fit SED of SGP 8884
        using templates including nebular emission (black lines and
        red symbols showing the synthetic fluxes) and standard
        spectral templates (magenta dashed line) for solar
        metallicity. For the other metallicities, the fits are very
        similar.  The upper limits (1\,$\sigma$ values) are also
        included in the SED fits.}
     \label{figure:sed_sgp}
   \end{figure}

\begin{figure}
   \centering
   \includegraphics[width=9cm]{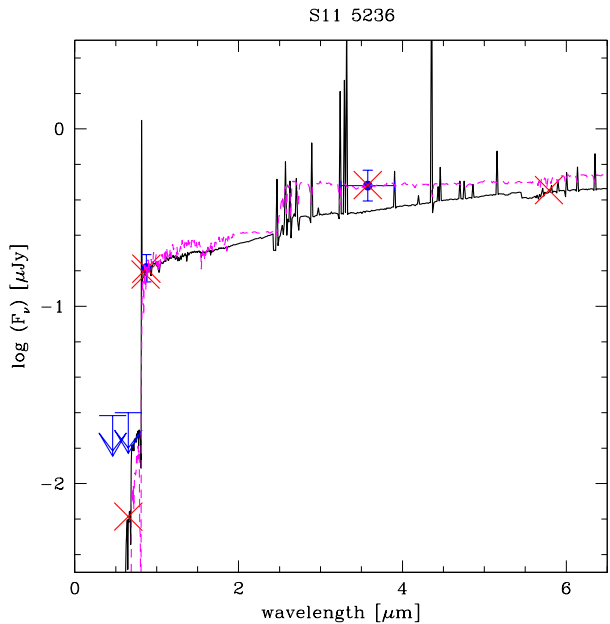}
      \caption{Observed and best-fit SED of S11 5236 using the same 
        symbols as in Fig.\ \protect\ref{figure:sed_sgp}.
        A constant SFR has been imposed for the fit with nebular lines
        to assure a sufficiently strong \lya\ emission (cf.\ text).
}
     \label{figure:sed_s11}
   \end{figure}

The broadband and spectroscopic measurements of the SED, summarised in
Table \ref{table:flux}, and the upper limits for the B and R bands
from \citet{Westra06} are used to constrain the SED fits.  Given the
relatively small number of observational constraints it is not
surprising that good SED fits are easily obtained (see
Figs.\ \ref{figure:sed_sgp} and \ref{figure:sed_s11}). We do not use
the strength of \lya that is derived from the modelling of the line to
constrain the fits directly, as the uncertainty in the intrinsic
strength of \lya is too large. For S11 5236, however, most models with
variable star-formation histories do not predict a strong enough
\lya\ emission. Therefore we have also examined fits imposing a
constant SFR.

The main qualitative difference between the two best-fit solutions
obtained with templates including nebular emission is obviously the
importance of the emission lines in the rest-frame optical, which is
predicted to be stronger for SGP~8884 than for S11~5236. The reason is
the non-detection at 5.8\,\mic in SGP~8884 with a flux below that at
3.6\,\mic, which drives the solution to a younger age, hence stronger
emission lines. For SGP~8884, approximately two-thirds of the
3.6\,\mic flux comes from line emission.

The probability distribution function (pdf) for the age, attenuation,
stellar mass, SFR, and specific SFR ($=SFR/\mstar$) for the two objects is shown
in Fig.\ \ref{figure:pdf} (black and red lines). The effect of imposing
a constant SFR is shown by comparing the solid and dashed lines.
As can be seen from these figures, the physical parameters of these
two objects span a relatively broad range of values.
In age and \mstar\ the two objects are quite distinct
(S11 5236 yielding older ages and a higher masses), reflected
by the two separate peaks in the pdfs.
For S11 5236 the pdf is skewed toward low {\em current} SFR values, as
the model favours short star-formation timescales and ``old'' ages.
For these cases, however, only weak \lya\ emission is expected, in
contrast with the observations. Assuming a long star-formation
timescale this shortcoming can be avoided; this situation is
e.g.\ illustrated by models with constant SFR (red dashed curves), in
which case the SFR and extinction are found to be somewhat larger, and
the stellar mass somewhat lower for S11 5236. The fact that short
  star-formation timescales are favoured (for both objects, but more
  so for S11~5236) also explains the relatively small overlap between
  the pdfs computed for variable $\tau$ models (including the case of
  constant SFR) and the constant SFR case.

For SGP 8884, the relatively high SFR and low mass implies a large
specific SFR (sSFR$\ga$ 100 Gyr$^{-1}$) if taken at face value.  In
any case, relatively large uncertainties on the physical parameters
remain, in particular given the small number of observational
constraints.

We now consider for comparison three other $z=5.7$ LAE
for which more photometric constraints are available, and discuss
the physical properties of these five LAEs together.

\subsection{Comparison with SED modelling of other $z=5.7$ LAE}
\begin{figure}
   \centering
   \includegraphics[width=9cm]{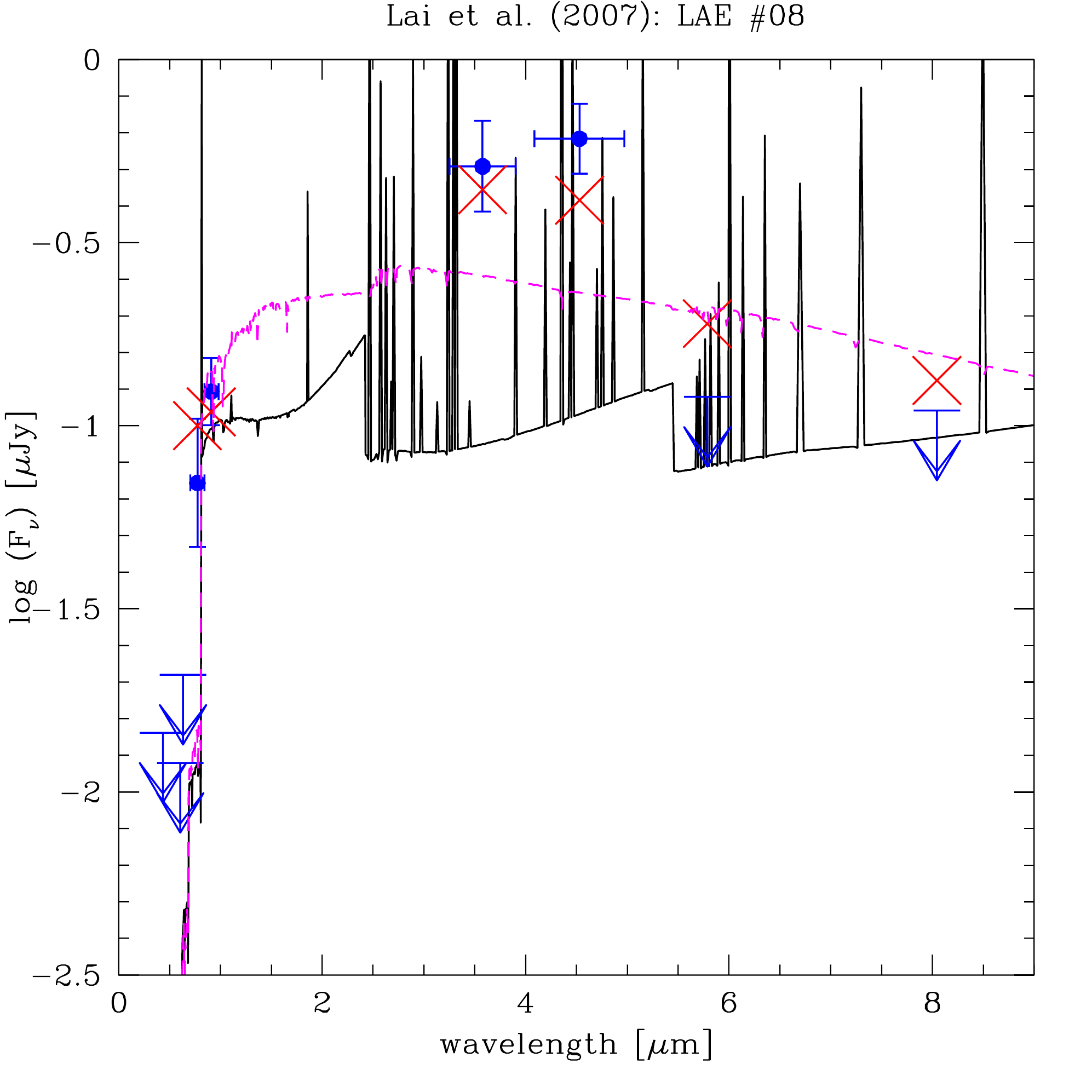}
       \caption{Observed (blue symbols) and best-fit SED of object
         \#08 from \citet{Lai07a}
        using templates including nebular emission (black lines and
        red symbols showing the synthetic fluxes) and standard 
        spectral templates (magenta dashed line). The upper limits 
        (1\,$\sigma$ values) are also included in the SED fits.}
     \label{figure:sed_lai8}
   \end{figure}

The three LAEs analysed earlier by \citet{Lai07a} are probably
still the best comparison sample for our two 2 bright LAEs.
These objects were also selected by their narrow-band excess,
then spectroscopically confirmed, and finally selected for 
being detected in the IRAC 3.6 and 4.5\,\mic images.
We have taken the published photometry from \citet{Lai07a} and
we have reanalysed these objects using the same code and assumptions
described above\footnote{We have noted that for these objects the pdf of the 
physical parameters does not change significantly if we allow for 
variable star-formation histories or assume constant SFR.}.
As an example, we show the best-fit SED of LAE \#08 in Fig.\
\ref{figure:sed_lai8}. Clearly the effect of nebular emission
(both lines and continua) on the best-fit SED are readily visible.
For all three objects, fits with nebular emission yield better
fits (lower $\chi^2$) than with standard templates.

\begin{figure*}
   \centering{   
     \includegraphics[width=7cm]{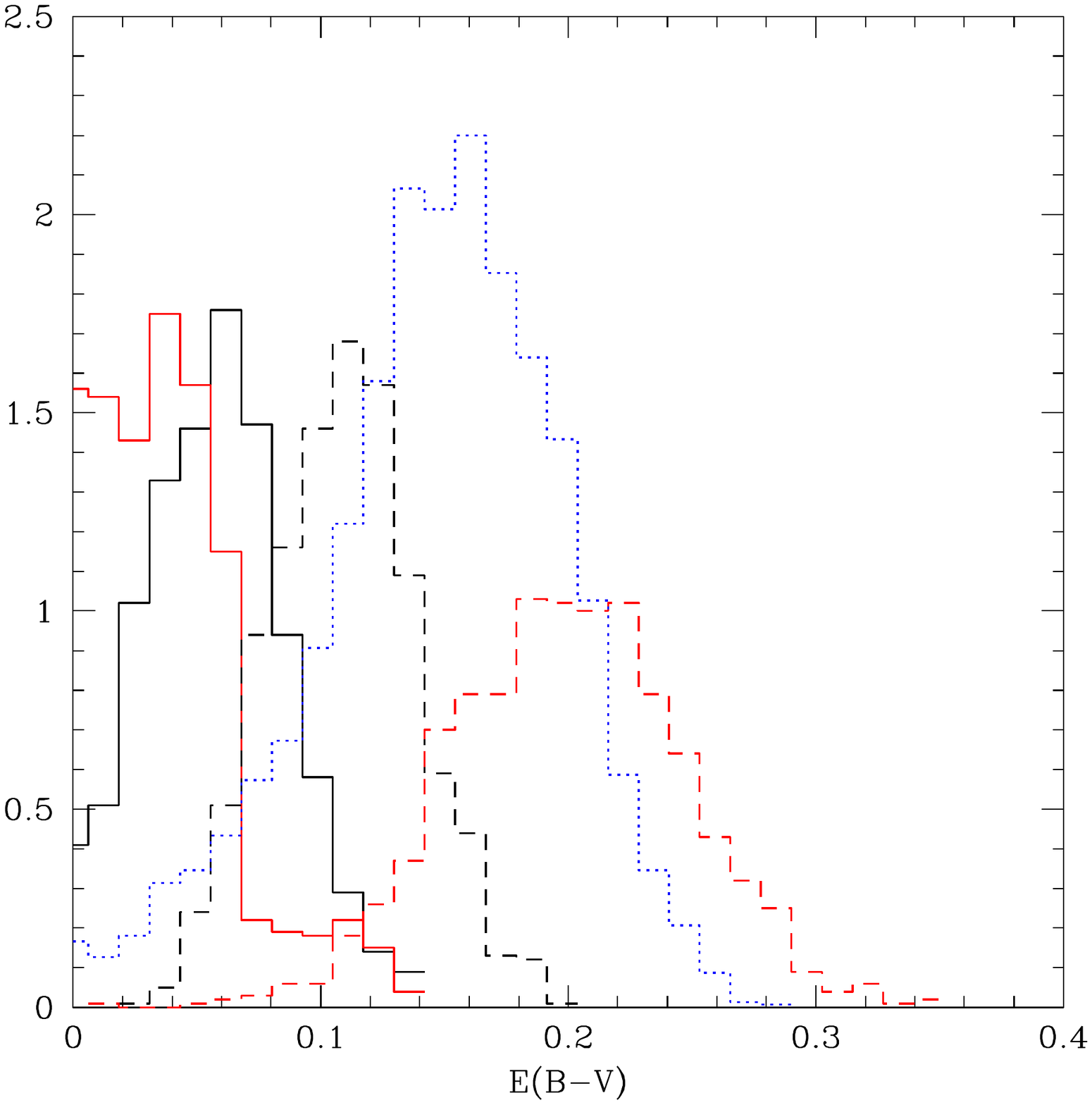}
     \includegraphics[width=7cm]{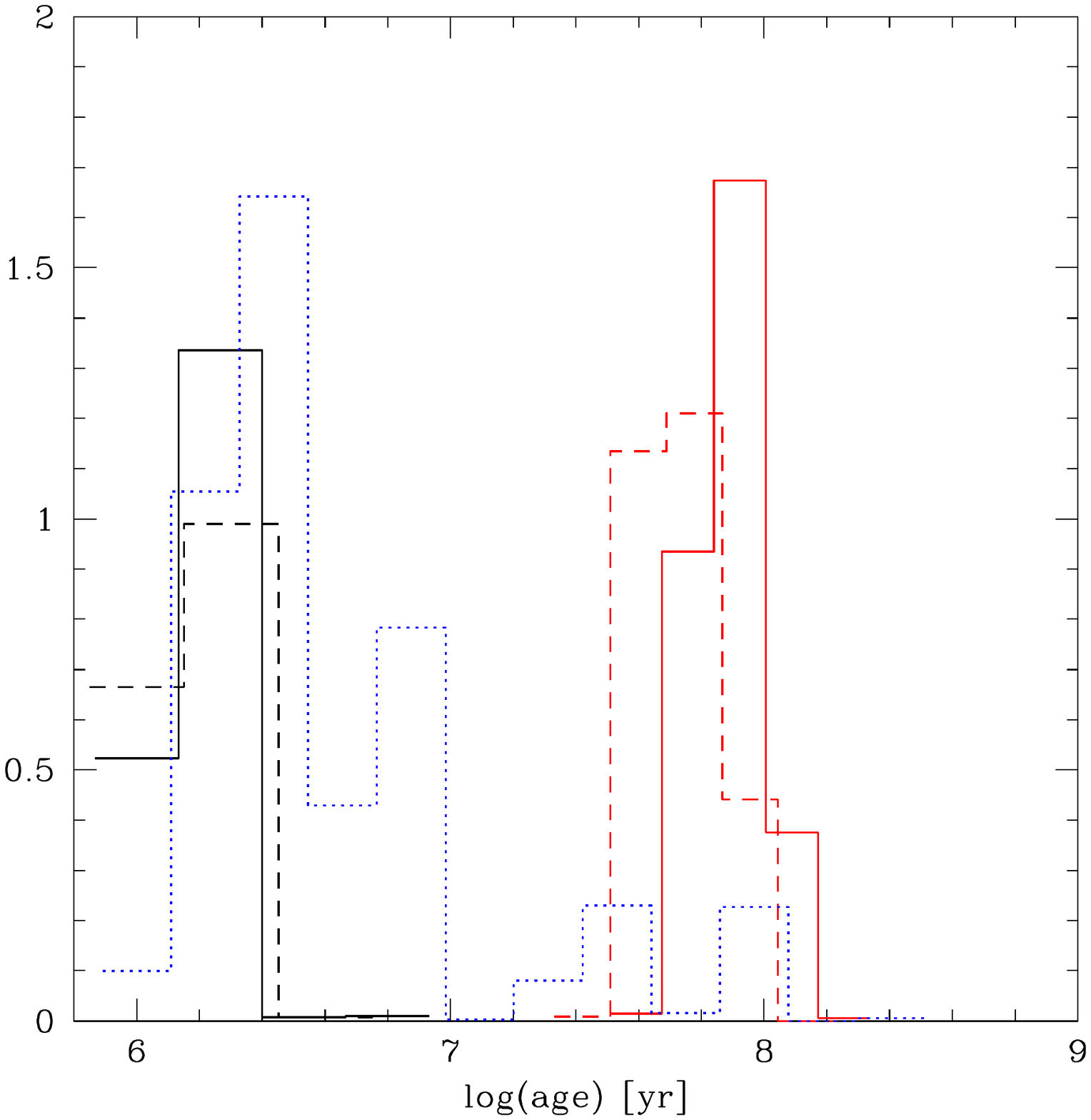}}
   \vspace{-2cm}
   \centering{
     \includegraphics[width=7cm]{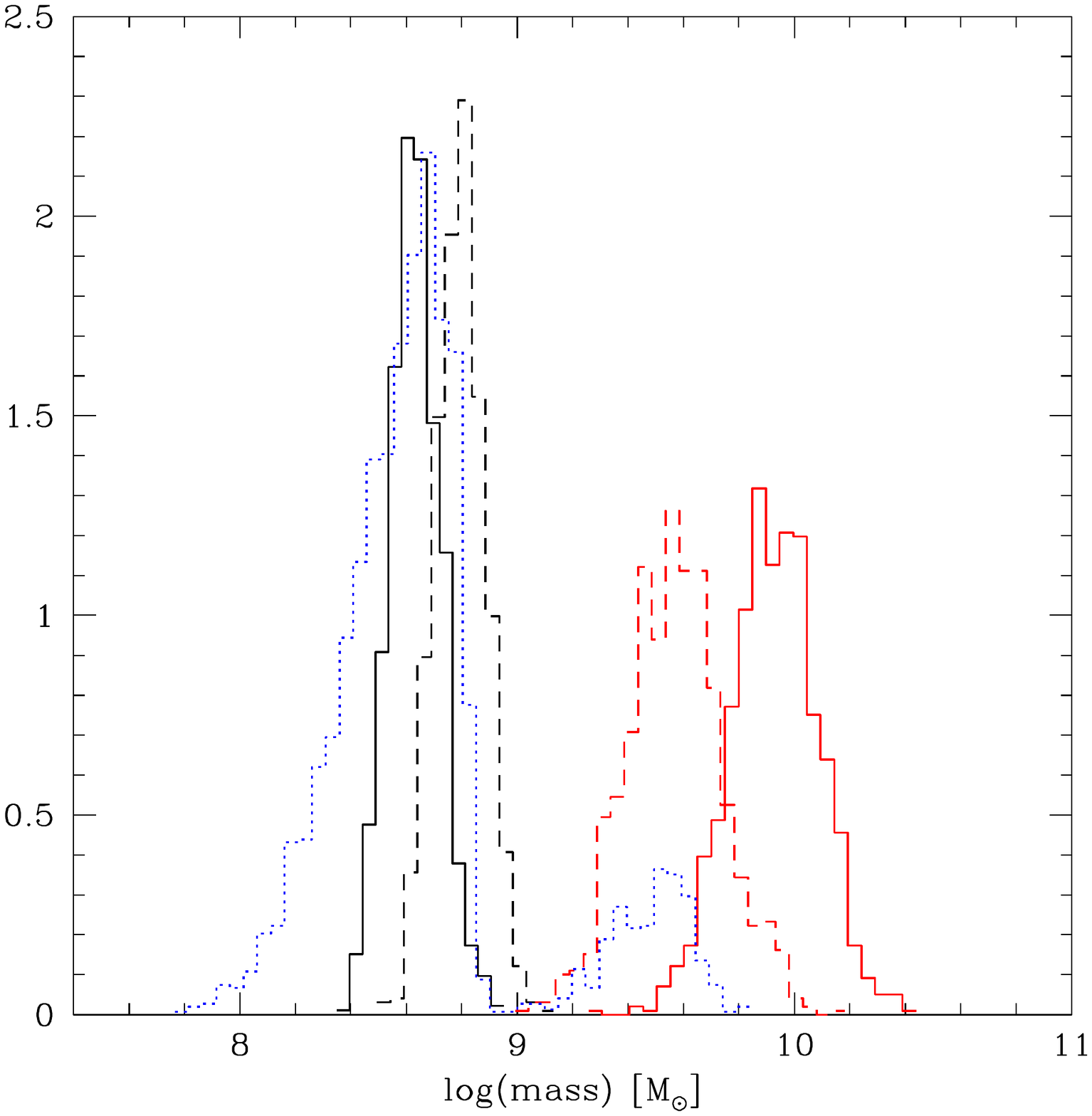}
     \includegraphics[width=7cm]{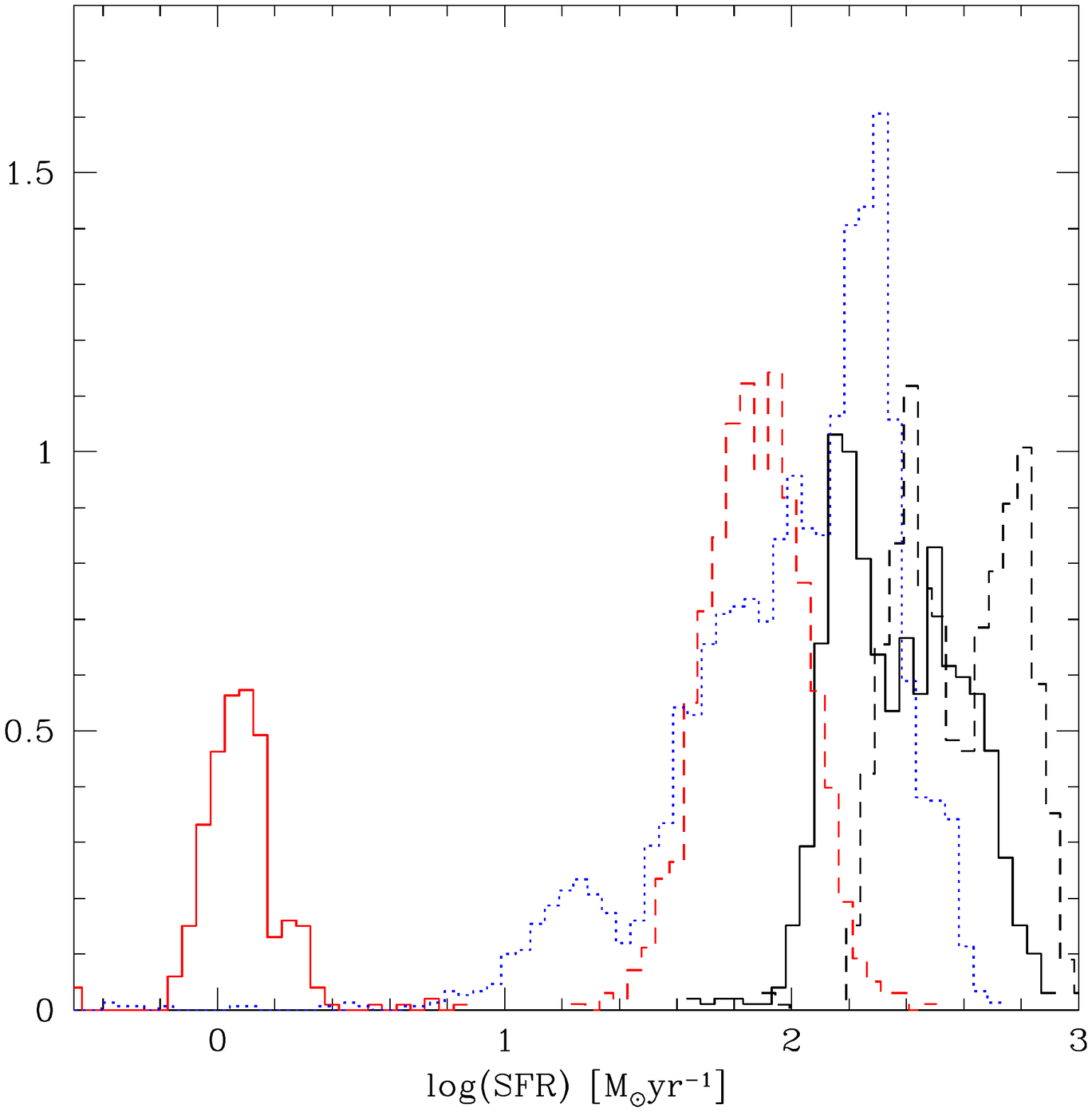}}
   \vspace{-2cm}
   \centering{
     \includegraphics[width=7cm]{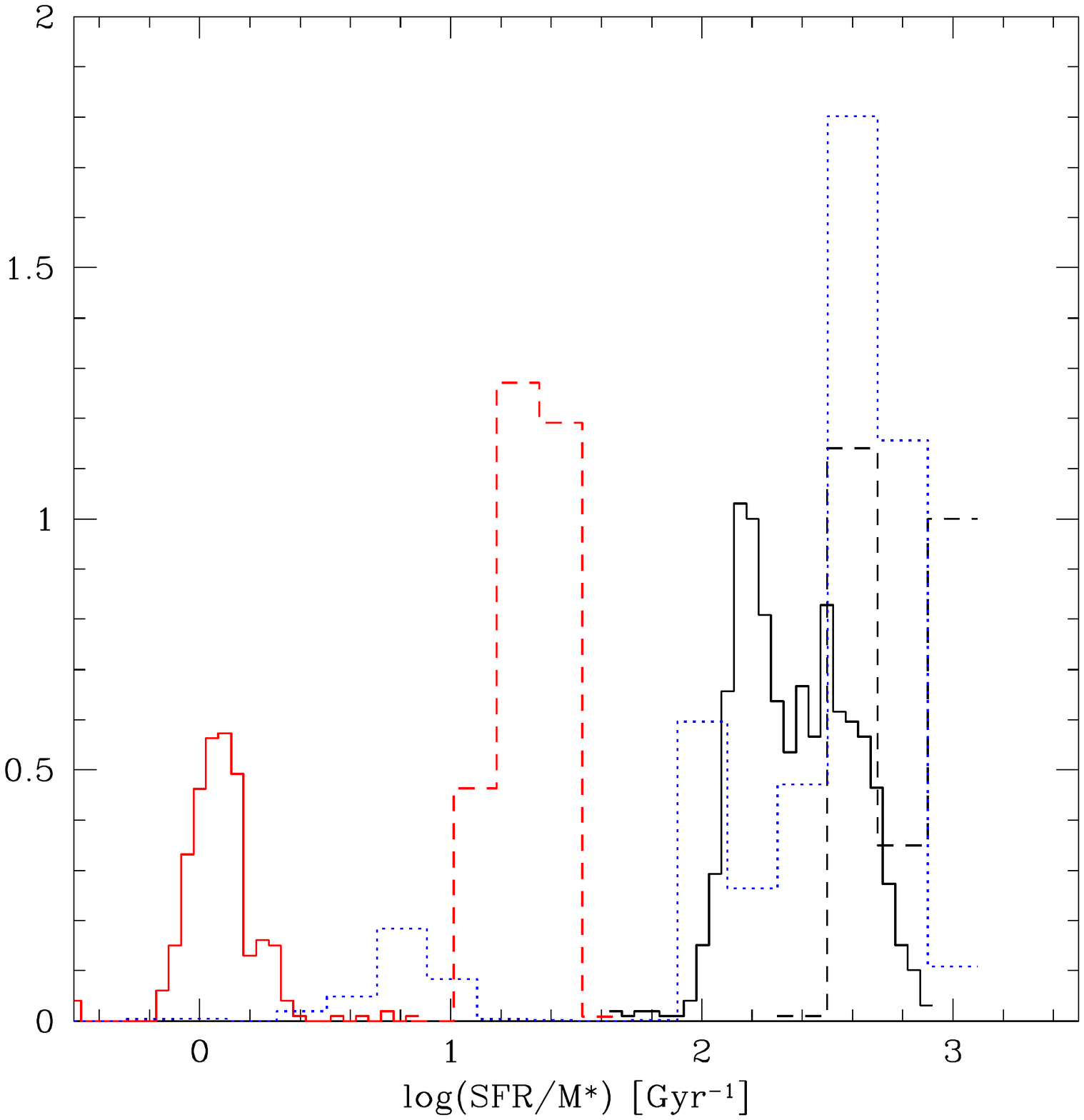}}
   \vspace{-2cm}
   \caption{Probability distribution functions (pdf) for the derived
     physical parameters of the two LAEs SGP 8884 (black) and S11 5236
     (red lines) and the 3 $z=5.7$ LAEs from \citet{Lai07a} (blue
     dotted lines). For our objects two pdfs are shown, assuming
     variable star-formation histories (solid lines) or constant SFR
     (dashed); the former underpredicts the observed \lya\ equivalent
     width of S11 5236.  For Lai's objects the pdf considers variable
     star-formation histories.  The area under the pdfs are
       normalised to a common value for the Lai et al. sample and for
       our objects taken together.  Note that for S11~5236 the pdf
       also extends to lower SFR than shown on the figure. {\bf Top
       left:} Attenuation \ebv\ derived for the \citet{Calz01} law.
     {\bf Top right:} Stellar age. {\bf Middle left:} stellar mass.
     {\bf Middle right:} instantaneous star-formation rate.  {\bf
       Bottom:} specific star-formation rate.  }
     \label{figure:pdf}
   \end{figure*}

As already illustrated in \citet{Lai07a} the SEDs of their other objects are 
very similar, both in relative and absolute fluxes. 
Therefore it is not surprising to find a simpler
probability distribution function for the physical parameters
of Lai's objects (see blue dotted lines in Fig.\ \ref{figure:pdf}) than
the more bimodal pdf for SGP~8884 and S11~5236.
We find that the properties of SGP~8884 resemble more the
typical values of the LAEs from Lai's sample, e.g.\ the SED, 
age, mass, SFR, and also \ebv.

Clearly do we find evidence for non-zero dust attenuation
in all 5 high-$z$ LAEs, as already found earlier 
by \citet{Lai07a} for their objects and e.g.\ by \citet{SP05}
for other $z \ga$ 6 LAEs. For example,
for Lai's sample we find $\ebv \approx 0.15 \pm 0.05$ and
for our two objects  $\ebv \approx$ 0.00--0.11 within 68\% confidence.

For the average stellar ages we do find a wide range of acceptable
values (see Fig.\ \ref{figure:pdf}). Overall, relatively young ages 
($\la$ 50--100 Myr) are however favoured, with a median slightly below 
10 Myr for Lai's sample.
Given these relatively young ages, the results from our SED fits
are quite insensitive to the exact assumed SF history.

The derived stellar masses also span a relatively large range,
typically between few times $10^8$ to few times $10^{10}$ \msun, the
largest masses being found for S11~5236.  For Lai's sample we find a
median stellar mass of $\sim 6 \times 10^8$ \msun, with 68\%
confidence limits between 2.2 and 6.2 $\times 10^8$ \msun.  Compared
to the analysis of \citet{Lai07a} our masses are significantly lower,
by a factor 10 approximately. The difference is due to several effects
in the SED fits: primarily the non-inclusion of nebular emission
(which dominates the flux at 3.6\,\mic and explains a factor $\sim$
3--4 difference), and differences in the absolute scaling of the
best-fit SEDs resulting from their peculiar treatment of the
non-detections in the 5.8 and 8.0\,\mic bands ($\sim$ factor 2.5).
Our masses and other fit parameters should be more reliable as those
from \citet{Lai07a} since our models include the effects of nebular
emission, which must by selection be particularly strong for LAEs.

The derived pdf for the SFR (Fig.\ \ref{figure:pdf}, bottom right)
shows again a rather large spread. For the objects from \citet{Lai07a} we find
SFR= $140^{+100}_{-105}$ \msunyr\ (68\% confidence interval), for our two
objects a more complex distribution. 
The ``standard'' \lya\ (UV) calibrations\footnote{See e.g.\ \citet{Kennicutt98a}
and \citet{Ajiki03} for \lya.}
 yield SFR $\approx$ 23--30 (20) \msunyr\
for our two objects, without extinction corrections. 
Except for the case of variable star-formation history for S11 5236
(cf.\ above) the (instantaneous) SFR derived from
the SED fits are higher due to non-zero extinction and due
to ages younger than the timescale ($\sim$ 100 Myr) assumed for the
UV calibration.

Finally it is interesting to note the large specific star-formation
rates (sSFR $\ga$ 10-100 Gyr$^{-1}$) found in general for the objects analysed here,
except for the fits using variable star-formation history models for S11 5236.
Below we discuss the sSFR values in comparison with other data.

\subsection{Comparisons with other data from the literature}
The values of the extinction and stellar mass derived for the bulk of the
LAEs (except maybe S11 5236) agree well with those found from a
larger sample of $z \approx$ 6--8 LBGs by \citet{SdB10}.
The specific SFR (SFR$/\mstar$) of the LAEs appears, however, somewhat
higher than average.

\citet{Ono10} have constructed the stacked SED of a large sample of 
LAEs at $z=5.7$ and 6.5, which are undetected at 3.6\,\mic by IRAC.
They have determined the average physical properties
from SED fits using models very similar to ours, including also
nebular emission. For their sample, showing a median \lya\ luminosity
of $(3.9 \pm 0.27) \times 10^{42}$ erg s$^{-1}$, they find a low stellar
mass (\mstar $\sim 10^{7.5}$ \msun), and no extinction.
These differences with our estimates are most naturally explained if on 
average more massive galaxies suffer from more extinction,
as well known from lower redshift \citep[e.g.,][]{Buat05,Buat08,Burgarella07,
Daddi07,Reddy06,Reddy08}
and also suggested to hold at the highest redshifts from observations
\citep{SdB10} and simulations \citep{Dayal10b}.
If \lya\ does not suffer much more strongly from extinction, it
may then also be expected, that the brightest \lya\ emitters at high-$z$
are also among the most massive galaxies, at least on average.
This is e.g.~supported by the recent semi-analytical models of 
\citet{Garel11a}, which use the results from \lya\ radiation
transfer calculations to predict the \lya\ escape fraction
depending on each galaxy's properties.

We also find that our median \ebv\ value ($\ebv \approx 0.15 \pm 0.05$)
determined for the LAE sample of \citet{Lai07a} is in very good 
agreement with the average continuum attenuation $f_c=0.23$ predicted for LAEs
at $z=5.7$ by the models of \citet{Dayal10a}.
It is also basically identical to that adopted in the cosmological model 
of \citet{Nagamine10a}.
However, the above \ebv\ value is larger than a ``cosmological average''
of the continuum attenuation at $z=5.7$ estimated by \citet{Hayes11a}.
Again, this is probably not surprising since our extinction value has been 
derived from relatively bright \lya\ emitters, which are also detected in
the IRAC bands. How our median values thus relate to other averages
is not trivial.

In any case, our analysis of 5 of the brightest LAEs at $z \sim 5.7$ 
confirms earlier findings that some of these very redshift galaxies
should contain some dust, and hence also be detectable in the 
infrared/sub-mm domain in the very near future \citep{SP05,Boone07,
Lai07a,Finkelstein09,Dayal10a,Dayal10b}.

Finally, we note that the specific star-formation rate sSFR $\ga$ 10-100 Gyr$^{-1}$ 
obtained here for the LAEs is significantly larger than the typical value 
of $\sim$ 2--3 Gyr$^{-1}$ found for LBGs by other authors at $z \sim 2$ and at 
redshifts up to $z>6$ \citep[see e.g.][]{Daddi07,Stark09,Gonzalez10,McLure11}. 
However, some LBGs at $z \sim 2-3$ \citep{erb06b,erb06c,yoshikawa10,DZ11},
some LAEs at  $z \sim 3-4$ \citep{Ono10},
and some LBGs at $z \ga 4$ \citep{SdB10,debarros11} also show comparably
high sSFR (and a large scatter)
Compared to recent semianalytic and hydrodynamic galaxy formation models
a low sSFR $\sim$ 2--3 Gyr$^{-1}$ constant with redshift above $z \ga 2$ 
seems problematic
\citep[cf.][]{Drory08,Nagamine10a,Bouche10,Khochfar11,Weinmann11}.
However, the high values indicated here appear larger than predicted
by these models, although they generally predict an increase of sSFR 
with decreasing galaxy mass, which might qualitatively explain why
the LAEs analyses here have high sSFR.
More in-depth SED studies of well constrained LAEs and LBGs at different 
redshift and independent measurements of SFR and mass will be needed to 
clarify these issues further.
  
\subsection{Dust extinction}

The modelling of the SED and the modelling of the \lya line provide
independent estimates of the amount of absorption from dust. While the
allowed values from the modelling of the \lya line are relatively
unconstrained, the estimates of the amount of absorption derived in
this way are generally different to the estimates derived from the
modelling of the SED. In particular, the estimate of the amount of
dust absorption in SGP~8884 is a factor of 2 to 3 higher than that
obtained from the SED fit. The difference has been noted in other LAEs
\citep{Verhamme08,Vanzella10a} and may be related to the difference
that has been seen local star-forming galaxies \citep{Calz01}, since one measure
is based on the fit to \lya, which is entirely nebular emission, and
the other is based on the fit to the SED, which is mixture of stellar
light, nebular emission and nebular continuum.  Alternatively, this
apparent difference may also be related to the simplified model
assumptions in the radiation transfer code (e.g.~simplified geometry)
or in the SED fits.

The \lya fits to both SGP~8884 and S11~5236 also result in unusually
low gas-to-extinction ratios, with values around
$N_\mathrm{H}/E(B-V)\sim 3-6\times
10^{19}\,\mathrm{cm^{-2}\,mag^{-1}}$ for the models that exclude the
effects of the IGM. Compared to the mean ratio in the Galaxy,
$N_\mathrm{H}/E(B-V)= 5.8\times 10^{21}\,\mathrm{cm^{-2}\,mag^{-1}}$,
\citep{Bohlin78a} the gas-to-extinction ratios in SGP~8884 and
S11~5236 are about 100 times lower. The values are also lower than
those estimated for other LAEs
\citep{Verhamme08,Vanzella10a}. However, the range allowed by the fits
is also large. For example, the 68.3\% confidence interval on the extinction
allows values as high $N_\mathrm{H}/E(B-V)=4\times
10^{20}\,\mathrm{cm^{-2}\,mag^{-1}}$ for SGP~8884, which is still
lower than that measured in the Galaxy, but similar to the values
estimated for other LAEs.  As already mentioned, it remains open
whether these results can be taken at face value or whether they are
due to simplifying assumptions in our model of modelling of the
\lya\ line.



\section{Summary}

We obtained deep optical and near-IR spectra and Spitzer IR imaging of
two of the most luminous LAEs known at $z=5.7$ and we use these data to
constrain the properties of these galaxies.

The continuum red-ward of \lya is clearly detected in both objects,
allowing us to derive reasonably precise estimates of the rest-frame
equivalent widths. For both objects we find equivalent widths
that are around $160\,\mathrm{\AA}$ with uncertainties of 10 and 20\%
for SGP~8884 and S11~5236, respectively.  

We used an analytic model and a 3-D Monte Carlo radiation transfer code
with two treatments for how the IGM affects the line to model the line
profile. All models, accurately model the blue edge of the line and
the continuum out to velocities of $\ga$ 1000 \kms\ blue-ward of
\lya. At a resolution of $5000$ and with the present S/N, we cannot
distinguish the three models.  Consequently, from the blue side of the
\lya line we do not find observational proof that \lya\
of our two LAE is affected by the IGM at these redshifts, although the
continuum is found to be depressed over a larger wavelength range
blue-ward of \lya.

However, neither the analytic model nor the model that excludes the
IGM from affecting the line profile adequately model the red wing of
the line in either LAE. Only the radiation transfer model that
includes a simple prescription for how the IGM affects the line
adequately models the red wing in both LAEs.  While this may be used
as evidence to support the idea that \lya is being affected by the IGM
at these redshifts, it is not yet conclusive, as the distribution of
dust and neutral hydrogen in the ISM in these LAEs is likely to be
more complex than assumed in the models used here.

The radiation transfer models predict that the intrinsic equivalent
width is about double the observed one, about 300\,$\mathrm{\AA}$
which is at the upper end of the range allowed for a young, moderately
metal-poor star-forming galaxy. This is independent of how we have
treated the IGM. However, uncertainties are currently large and values
as high as 700\,$\mathrm{\AA}$, which in the realm expected for a
Population III burst, are allowed.

Both LAEs were observed and detected with IRAC on the {\it Spitzer Space Telescope}. We
combined the IRAC photometry with a measurement of the continuum at
1300$\,\mathrm{\AA}$ and ground based data to constrain the SED of
these two objects and compared them to 3 other LAEs at $z=5.7$ from
the literature \citep{Lai07a}. The SEDs of these 5 LAEs have been
analysed with state-of-the-art SED models including the effects of
nebular emission \citep[cf.][]{SdB09,SdB10}.  In terms of age and
mass, our two LAEs are quite distinct from each-other. S11~5236
appears older and more massive than SGP~8884.  Four of the five LAEs
analysed here show masses of the order of $\sim 5 \times 10^8$ \msun, lower
than previous estimates of Lai et al. We find evidence for the presence of
some dust in all objects, and indications for fairly high specific
star-formation rates ($\ga$ 10--100 Gyr$^{-1}$).

Nebular line emission, principally from [OIII]\,$\lambda\lambda$
4959,5007 doublet and H$\beta$, makes a significant contribution to
the flux in the IRAC 3.6\,\mic filter. For SGP~8884 and the three
objects in \citet{Lai07a}, it dominates.  Dusty, and rich in emission lines,
these objects will be prime targets for the next generation of
extremely large telescopes, JWST and ALMA.

\section*{Acknowledgements}

C. Lidman wishes to acknowledge the support of the
Oskar Klein Centre at the University of Stockholm and the support of
the Australian Research Council (ARC) through the ARC Future
Fellowship. DS and MH acknowledge support from the Swiss National Science
Foundation.



\end{document}